\documentclass[12]{article}

\usepackage{amsfonts, amsmath, amssymb}
\usepackage{etoolbox}
\usepackage[utf8]{inputenc} 
\usepackage{hyperref}  
\usepackage{url}            
\usepackage{booktabs}       
\usepackage{amsfonts}       
\usepackage{nicefrac}       
\usepackage{microtype}      
\usepackage{lipsum}
\usepackage{graphicx}
\usepackage{subcaption}
\usepackage{color} 
\usepackage{cite}
\usepackage{authblk}
\usepackage{tikz}
\usepackage{orcidlink}
\graphicspath{ {./imageskQM/} }

\title{Spectral design of confining optical structures by supersymmetry techniques}

\author[1$\dagger$]{\orcidlink{0009-0009-1532-8184} Samantha Cervantes-Neri} 
\author[2*]{\orcidlink{0000-0002-7051-6147} Zulema Gress-Mendoza}
\author[1,3$\S$]{\orcidlink{0000-0003-0321-3442} Daniel O-Campa}  
\author[1$\ddagger$]{ \orcidlink{0000-0002-2180-3895} Erik D\'iaz-Bautista}
\author[4$\star$]{\orcidlink{0000-0002-7133-0803} Sara Cruz y Cruz}

\affil[1]{\small Instituto Polit\'ecnico Nacional, ESFM, Av. Luis Enrique Erro S/N 07738 Ciudad de M\'exico, Mexico}
\affil[2]{\small \'Area Acad\'emica de Matem\'aticas y F\'isica, Universidad Aut\'onoma del Estado de Hidalgo, 42184, Pachuca, Hidalgo, Mexico}
\affil[3]{\small División de Ingeniería Electromecánica, Instituto Tecnológico Superior del Occidente del Estado de Hidalgo, 42700, Mixquiahuala de Juárez, Hidalgo, Mexico}
\affil[4]{\small Instituto Polit\'ecnico Nacional, UPIITA, Av. Instituto Polit\'ecnico Nacional 2580 07340 Ciudad de M\'exico, Mexico}

\date{\small e-mail: $^\dagger$cervantesneri.samantha@gmail.com, $^*$zulema\_gress@uaeh.edu.mx, $^\S$dortiz@itsoeh.edu.mx, $^\ddagger$ediazba@ipn.mx, $^\star$sgcruzc@ipn.mx}
\begin{document}
\maketitle

\begin{abstract}
In this work, we apply supersymmetric transformations to an optical medium with quadratic index distributions on both transversal coordinates defining an elliptic profile. In particular, we implement first- and second-order  transformations to construct new multi-parametric gradient index patterns that may host one or two subsidiary guided modes with a wide range of spectral properties and transversal intensity and phase distributions. These transformations may be continued up to an arbitrary order enabling the possibility to control the spectral and propagation properties of light beams.
\end{abstract}


\section{Introduction}
\label{Intro}

The propagation of light beams in inhomogeneous materials is a fundamental phenomenon in optics, with applications ranging from optical fibers to advanced photonic devices \cite{Gom02}. Yet, the stocktaking of exactly solvable models in this field is not so large. The importance of analytical beam-like solutions of the wave equation lies in the fact that their behavior in propagation and transformation processes can be exactly predicted enabling the design of flexible and precise protocols of light manipulation. In this sense, the profound correspondence between wave optics and quantum mechanics, in which refractive-index landscapes to be interpreted as optical analogues of quantum potentials \cite{Mar97,Dra99,Dra02,Lon09,Cru15,Cru15a}, plays a significant role. This formal equivalence provides with new methods, originally developed to construct exactly solvable models in quantum systems, to be applied in classical optics. We refer to operatorial methods, factorization and spectral engineering \cite{Lon09,Glo69,Sto81,Nie93,Cru17,Cru19,Lon18,Boc22,Cru23}. A particularly powerful technique comprises supersymmetric or Darboux transfomations \cite{Chu94,Mir13,Hei14}. In this approach, two or more Hamiltonians $H^{(j)}$, $j=0,1,2,\ldots$, sharing a part of their spectra, are intertwined through a set of intertwining operators \cite{Mie84,Jun98,Mie00,Coo01,Mie04,Suk04,Fer04,Gan10,Ber14,Var23}. The consequence is that the discrete spectra as well as the corresponding space of solutions of each Hamiltonian, can be completely determined if the complete set of solutions of one departing eigenvalue problem, say $H^{(0)} \psi^{(0)} = \varepsilon^{(0)} \psi^{(0)}$, is already known. Based on the seminal work of Chumakov and Wolf \cite{Chu94}, it is now understood that the Helmholtz equation admits supersymmetric transformations, allowing the construction of new exactly solvable models and the spectral design of optical architectures.

The classical-quantum analogy provides both, conceptual and practical advantages in different contexts. From the conceptual point of view, the factorization scheme allows the recasting of optical modes, propagation constants and radiation responses in terms of operator algebras and group theoretical methods. On the other hand, from the practical perspective, the supersymmetric transformations offer a systematic mechanism for generating refractive-index profiles with prescribed modal properties: partner structures may be designed to share the same number of guided modes, to remove or add specific ones, or to redistribute modal content in controlled fashion. Recent developments in integrated photonics and waveguide arrays have demonstrated these ideas are experimentally achievable, enabling selective mode filtering, robust mode conversion, and isospectral scattering structures derived from supersymmetric transformations \cite{Mir13,Hei14,Mac18,Hua22}.

The resulting framework naturally bridges classical optics and quantum-inspired design principles. Unlike optical engineering relying on geometric and wave-based schemes, supersymmetry techniques provide algebraic tools, such as Hilbert spaces, underlying algebras and coherent states, offering new routes for constructing structured light and engineered mode sets. Moreover, supersymmetry-based methods complement other symmetry-driven approaches, such as PT symmetry \cite{Boc22}, transformation optics, and topological photonics, thus expanding the palette of available design strategies. Extensions of supersymmetry to the time domain further emphasize its versatility and potential for dynamic control of optical fields \cite{Zel17,Raz19,Con19,Cru20,Gar20}.

In this work we revisit the wave optics-quantum mechanics correspondence, following the formalism presented in \cite{Cru17,Gre17,Gre19,Cru19,Cru23},  in order to apply supersymmetric techniques in the spectral design of new GRIN  (gradient refractive index) optical structures and shaping of guided modes. First, in Section \ref{SusyQM} we present a brief overview of the supersymmetric transformations of first and higher order. Next, in Section \ref{sec:headings} we introduce a class of GRIN wave model, in correspondence with the stationary Sch\"odinger eigenvalue problem. As an example, the guided (stationary) modes for the elliptical GRIN wave guide are determined, along with the spectrum of propagation constants. Section \ref{SusyO}, in turn, contains new supersymmetry generated optical structures together with some examples addressed to illustrate our formalism. Finally, Section \ref{Conc} present some conclusions. 

\section{Supersymmetry transformations of \emph{k}-th order}
\label{SusyQM}

\subsection{First order supersymmetric transformations}

Supersymmetry transformations constitute a mathematical tool that has been extensively used to generate exactly solvable models in quantum mechanics \cite{Mie04,Mie00,Mie84,Gan10,Coo01,Suk04}. In the supersymmetric approach, it is assumed that  two second-order, one-dimensional Hamiltonian operators of the form
\begin{equation}
\label{16}
H^{(j)}=-\frac12 \partial^2_{\rm x} +V^{(j)}({\rm x}), \quad j=0,1,
\end{equation}
are intertwined by an operator $L_1^+$ in such a way that
\begin{equation}
\label{intertwining}
H^{(1)}L_{1}^{+}=L_{1}^{+}H^{(0)}.
\end{equation}
In the equation \eqref{16}  $V^{(j)}$, $j=0,1$ are two-real valued potentials and we have adopted the short notation $\partial_\upsilon = \frac{\mathrm{d}}{\mathrm{d} \upsilon}$ and $\partial^2_\upsilon = \partial_\upsilon \cdot \partial_\upsilon$. The intertwining operator $L_1^+$ can be constructed as a first order differential one
\begin{equation}\label{ec17}
L_{1}^{+}= \frac{1}{\sqrt{2}}\left[-\partial_{\rm x} + W_{1}({\rm x})\right],
\end{equation}
with $W_{1}$ a real-valued function to be determined called superpotential. Also, it is assumed that the potential $V^{(0)}$ is well known, along with the whole set of solutions to the eigenvalue problem
\begin{equation}
\label{EigenEqH0}
H^{(0)} \psi^{(0)} = \varepsilon \psi^{(0)}, \qquad \varepsilon \in \mathbb{R}.
\end{equation}
The premisses \eqref{16}-\eqref{EigenEqH0} have a series of non-trivial consequences. In the first place, the substitution of \eqref{16} and \eqref{ec17} into \eqref{intertwining} leads to the set of equations
\begin{equation}\label{Riccati}
\partial_{\rm x} W_{1}({\rm x},\epsilon_1)+W_{1}^{2}({\rm x},\epsilon_1)=2\left[V^{(0)}({\rm x})-\epsilon_1\right],
\end{equation}
\begin{equation}\label{V+V-}
V^{(1)}({\rm x},\epsilon_1)=V^{(0)}({\rm x})- \partial_{\rm x} W_{1}({\rm x},\epsilon_1),
\end{equation}
with $\epsilon_1 \in \mathbb{R}$ an integration constant to be fixed. The equation \eqref{Riccati} is a Riccati equation that allows to determine the superpotential once the parameter $\epsilon_1$ is fixed, generating to a family of intertwining operators $L_1^+(\epsilon_1)$. In turn, the equation \eqref{V+V-} outlines the form of the corresponding family $V^{(1)}(\epsilon_1)$. Also, the set of equations  \eqref{Riccati}-\eqref{V+V-} allows to express the Hamiltonians $H^{(0,1)}$ in the factorized form
\begin{equation}\label{ec26}
H^{(0)}=L_{1}^{-}(\epsilon_1) L_{1}^{+}(\epsilon_1)+\epsilon_1, \qquad H^{(1)}(\epsilon_1)=L_{1}^{+}(\epsilon_1) L_{1}^{-}(\epsilon_1)+\epsilon_1.
\end{equation}
 where $L^-_1(\epsilon_1)$ is the adjoint intertwining operator 
\begin{equation}\label{ec25}
L_{1}^{-}(\epsilon_1)=\left(L_{1}^{+}(\epsilon_1)\right)^\dagger=\frac{1}{\sqrt{2}} \left[ \partial_{\rm x} + W_{1}({\rm x},\epsilon_1) \right].
\end{equation} 

\begin{figure}[ht]
\centering
\includegraphics[width=0.325\linewidth]{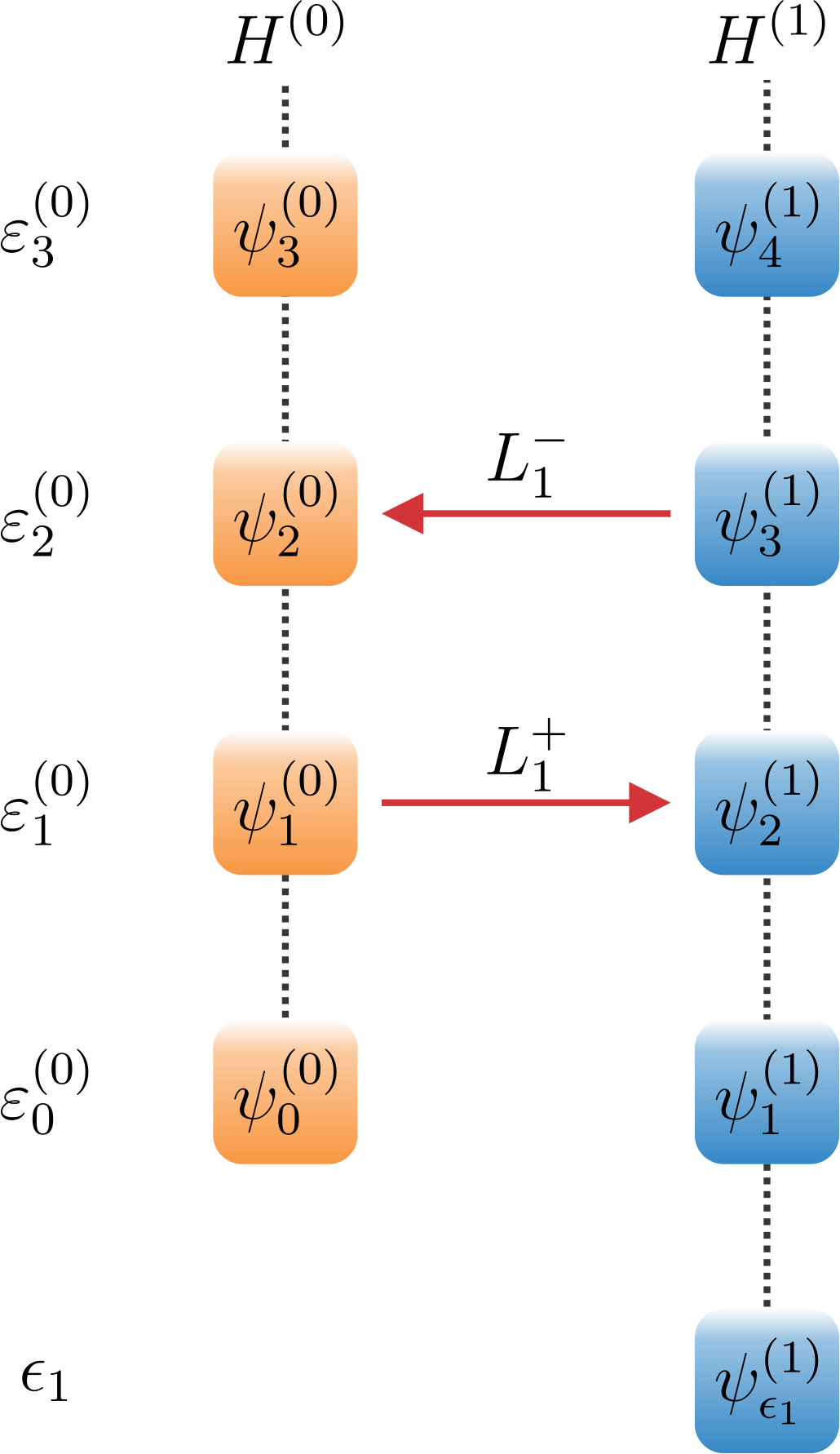}
\caption{Action of the intertwining operators $L_1^{\pm}$ on the eigenfuntions of $H^{(0,1)}$. In case that the selection of $\epsilon_1$ and the transformation function yields a square integrable function $\psi_{\epsilon_1}^{(1)}$, then the supersymmetric transformation deforms the departure potential $V^{(0)}$ in such a way that the arriving one $V^{(1)}$ hosts exactly one additional eigenstate associated to the eigenvalue $\varepsilon_0^{(1)} = \epsilon_1$.}
\label{fig:SUSYQM}
\end{figure}

This formalism enables the construction of new exactly solvable models, as supersymmetric transformations of very well known ones. The problem is reduced to the determination of the function $W_1(\rm x,\epsilon_1)$, associated to a particular selection of $V^{(0)}$ and $\epsilon_1$, by solving the Riccati equation \eqref{Riccati}.  Typically, this equation is linearized with the introduction of the change of variable
\begin{equation}\label{ec21}
W_{1}({\rm x},\epsilon_1) = \partial_{\rm x} \left( \ln u^{(0)}({\rm x},\epsilon_1) \right),
\end{equation}
where $u^{(0)}$, known as \emph{transformation} or \emph{seed function}, satisfies the departure eigenvalue problem
\begin{equation}
\label{seedfunction}
H^{(0)} u^{(0)}({\rm x},\epsilon_1) = \epsilon_1 \, u^{(0)}({\rm x},\epsilon_1),
\end{equation}
for which the complete sets of solutions for each value of $\epsilon_1$ are well known and characterized. In terms of $u^{(0)}$, the new potential $V^{(1)}$ turns into
\begin{equation}
\label{V1u}
V^{(1)}({\rm x},\epsilon_1) = V^{(0)}({\rm x}) - \partial^2_{\rm x} \ln u^{(0)}({\rm x},\epsilon_1).
\end{equation}
Clearly, the transformation function $u^{(0)}$ must be chosen to have no nodes on its domain in order to avoid singularities of $V^{(1)}$ that are not inherited from those of $V^{(0)}$. 
Furthermore, the exact solvability of the new model may be readily checked. Indeed, assuming that $\psi^{(0)}({\rm x},\varepsilon)$ is a solution of \eqref{EigenEqH0}, the intertwining relation \eqref{intertwining} implies that the function 
\begin{equation}
\label{psi1}
\psi^{(1)}({\rm x},\epsilon_1,\varepsilon) \propto L_1^+(\epsilon_1) \psi^{(0)}({\rm x},\varepsilon)
\end{equation} 
satisfies
\begin{equation}
\label{EigenEqH1}
H^{(1)}(\epsilon_1) \psi^{(1)}({\rm x},\epsilon_1,\varepsilon) = \varepsilon \psi^{(1)}({\rm x},\epsilon_1,\varepsilon).
\end{equation}
This fact is not limited to physically interpretable solutions. Yet, if $\varepsilon = \varepsilon^{(0)}_n$, $n = 0,1,2,\ldots$ belongs to the discrete spectrum of $H^{(0)}$, then $\psi^{(0)}({\rm x},\varepsilon^{(0)}_n) \equiv \psi^{(0)}_n({\rm x})$ can be chosen as square integrable. The corresponding function
$$
\psi^{(1)}({\rm x},\varepsilon_{n+1}^{(1)}) = \psi^{(1)}({\rm x},\varepsilon_{n}^{(0)}) = \psi^{(1)}_{n+1}({\rm x}) \propto L_1^+(\epsilon_1)  \psi^{(0)}_n({\rm x}), \qquad \varepsilon_{n+1}^{(1)} = \varepsilon_n^{(0)},
$$ 
is also square integrable \cite{Inf51}, so that $\varepsilon_{n+1}^{(1)} = \varepsilon_n^{(0)}$ do also belongs to the discrete spectrum of $H^{(1)}(\epsilon_1)$. 
In this way, if we suppose that $\psi^{(0)}_n$, $n=0,1,2,\ldots$ are normalized, then 
\begin{equation}\label{ec27}
\vert\vert \psi^{(1)}_{n+1}\vert\vert^2 \propto \vert\vert L_{1}^{+}(\epsilon_1)\psi_{n}^{(0)}\vert\vert^2=\varepsilon_{n}^{(0)}-\epsilon_1 \geq 0.
\end{equation}
This inequality is fulfilled whenever $\epsilon_1 \leq \varepsilon_{0}^{(0)}$, and provides a restriction on the values that the parameter $\epsilon_1$ may take. Therefore, the normalized eigenfunctions $\psi_{n+1}^{(1)}({\rm x})$ of the Hamiltonian $H^{(1)}(\epsilon_1)$ associated to the eigenvalue $\varepsilon_{n+1}^{(1)}$ are constructed as
\begin{equation}\label{ec28}
\psi_{n+1}^{(1)}({\rm x}) = \frac{L_{1}^{+}(\epsilon_1) \, \psi_{n}^{(0)}({\rm x})}{\sqrt{\varepsilon_{n}^{(0)} - \epsilon_1}}, \qquad \varepsilon_{n+1}^{(1)}=\varepsilon_{n}^{(0)}, \quad n = 0, 1, 2, \dots.
\end{equation}
It is worthwhile to point out that, even when the set $\mathcal{S}^{(0)} = \left\{\psi_n^{(0)}:~n=0,1,2,\ldots\right\}$ is complete in the Hilbert space $\mathcal H^{(0)}$ of square integrable solutions of \eqref{EigenEqH0}, the set $\left\{\psi_{n}^{(1)}:~n=1,2,3,\ldots\right\}$ is, in general, not complete in the space $\mathcal H^{(1)}$ of square integrable solutions of \eqref{EigenEqH1}. Indeed, assuming the function $\psi_{\epsilon_1}^{(1)}$ is orthogonal to each $\psi_{n}^{(1)}$, $n = 1,2,\dots$, we arrive to the condition $L_{1}^{-}(\epsilon_1) \psi_{\epsilon_1}^{(1)}(\rm x) = 0$. Notice that, according to \eqref{ec26}, this function is a solution of \eqref{EigenEqH1} with spectral parameter $\varepsilon = \epsilon_1$. Furthermore, as $L_1^{-}$ is a first order differential operator, we conclude that $H^{(1)}(\epsilon_1)$ may have, at most, one eigenfunction more than $H^{(0)}$ corresponding to the eigenvalue $\varepsilon_0^{(1)} = \epsilon$. Moreover, a direct calculation shows that this function is given by   
\begin{equation}\label{ec29}
\psi_{\epsilon_1}^{(1)}({\rm x}) = \mathcal{C}_1 \exp\left(-\int^{{\rm x}} W_{1}(t,\epsilon_1) \, \mathrm{d}t\right) = \frac{\mathcal{C}_1}{u^{(0)}({\rm x},\epsilon_1)},
\end{equation}
where $\mathcal{C}_1$ is a normalization constant and $u^{(0)}({\rm x},\epsilon_1)$ is the transformation function. In this way, if $\psi_{\epsilon_1}^{(1)}$ turns out to be square integrable, then the spectrum $\mathrm{Sp}\left(H^{(1)}\right) = \left\{\epsilon_1, \varepsilon^{(0)}_n:~n=0,1,2\ldots\right\}$ contains exactly one element more than $\mathrm{Sp}\left(H^{(0)}\right)$ and $\mathcal{S}^{(1)} = \left\{\psi^{(1)}_{\epsilon_1}, \psi_{n+1}^{(1)}: ~n=1,2,3,\ldots\right\}$ (see Figure \ref{fig:SUSYQM}).

\subsection{The iterative approach}

\subsubsection{Second order supersymmetric transformation}

Once we have constructed the exactly solvable Hamiltonian $H^{(1)}(\epsilon)$, we may iterate the supersymmetric transformation to construct a new one  \cite{Mie00,Fer04}.  To this end, let us fix $\epsilon = \epsilon_1$ and consider a new Hamiltonian  
$$
H^{(2)}(\epsilon_1,\epsilon_2) = -\frac12 \partial^2_{\rm x} + V^{(2)}({\rm x},\epsilon_1,\epsilon_2), \qquad \epsilon_2< \epsilon_1,
$$
verifying
\begin{equation}
\label{ec29.2}
H^{(2)}(\epsilon_1,\epsilon_2) L_2^+(\epsilon_1,\epsilon_2) = L_2^+(\epsilon_1,\epsilon_2) H^{(1)}(\epsilon_1).
\end{equation}
with the new intertwining operator
\begin{equation}\label{ec29.1}
    L_{2}^{+}(\epsilon_1,\epsilon_2)=\frac{1}{\sqrt{2}}\left[-\partial_{\rm x}+W_{2}({\rm x}.\epsilon_1,\epsilon_{2})\right]
\end{equation}
Equivalently to the first order case, we are led to the set of equations
\begin{equation}\label{ec29.3}
\partial_{\rm x} W_{2}({\rm x},\epsilon_1,\epsilon_{2}) + W_2^2({\rm x},\epsilon_2,\epsilon_2)=2\left[V^{(1)}({\rm x},\epsilon_1)-\epsilon_{2}\right], \qquad  V^{(2)}({\rm x},\epsilon_1,\epsilon_2)=V^{(1)}({\rm x},\epsilon_1)-\partial_x W_{2}({\rm x},\epsilon_1,\epsilon_{2}),
\end{equation}
that may be addressed introducing the transformation function $u^{(1)}(\epsilon_1,\epsilon_2)$ satisfying
\begin{equation}\label{ec29.4}
    W_{2}({\rm x},\epsilon_1,\epsilon_{2}) =\partial_{\rm x} \left( \ln u^{(1)}({\rm x},\epsilon_1,\epsilon_2) \right), \qquad H^{(1)}(\epsilon_1) u^{(1)}({\rm x},\epsilon_1,\epsilon_2) = \epsilon_2 u^{(1)}({\rm x},\epsilon_1,\epsilon_2).
\end{equation}
Notice that the transformation function $u^{(1)}(\epsilon_1,\epsilon_2)$ may be written in the form \eqref{psi1} for a function $u^{(0)}(\epsilon_2)$ such that $H^{(0)}u^{(0)}({\rm x},\epsilon_2) = \epsilon_2 u^{(0)}({\rm x},\epsilon_2)$. As $\epsilon_1\neq\epsilon_2$, then $u^{(0)}(\epsilon_1)$ and $u^{(0)}(\epsilon_2)$ are orthogonal each other. In this way 
 \begin{equation}\label{ec29.6}
     u^{(1)}({\rm x},\epsilon_1,\epsilon_2) \propto L_{1}^{+}(\epsilon_1)u^{(0)}({\rm x},\epsilon_2)\propto-\frac{\mathcal{W}\left(u^{(0)}(\epsilon_1),u^{(0)}(\epsilon_2)\right)}{u^{(0)}({\rm x},\epsilon_1)},
 \end{equation}
 where $\mathcal{W}(f,g)$ denotes the Wronskian of $f$ and $g$. This means that the problem to obtain the transformation function $u^{(1)}(\epsilon_1,\epsilon_2)$ defining the second order supersymmetric transformation is reduced to the fixing of two non-vanishing solutions of the departure eigenvalue equation  with $\varepsilon = \epsilon_1$ and $\varepsilon = \epsilon_2$, $\epsilon_2 < \epsilon_1$. 
 
\begin{figure}[ht]
    \centering
    \includegraphics[width=0.65\linewidth]{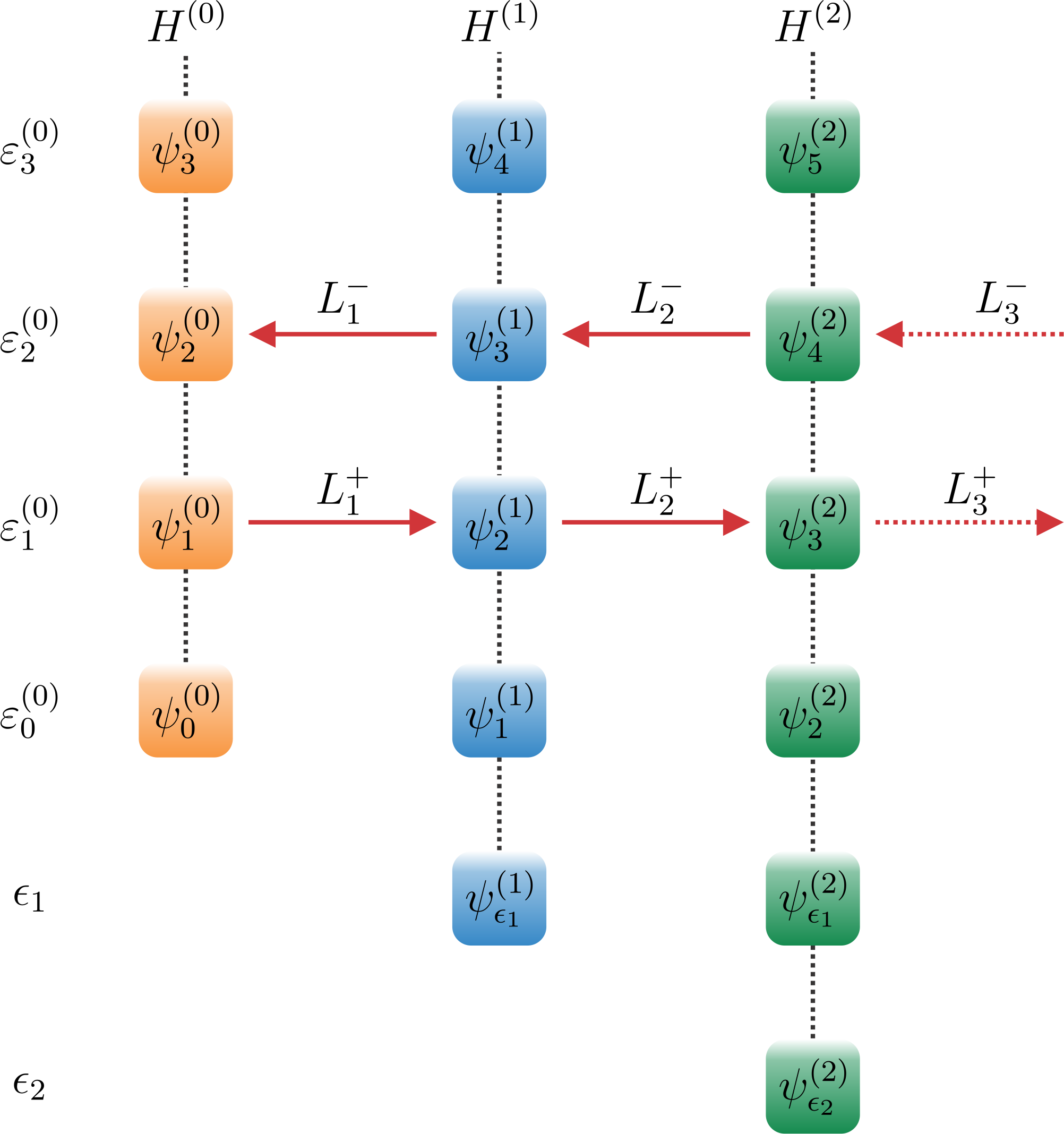}
\caption{Iterative action of the intertwining operators $L_j^{\pm}$, $j=1,2,\ldots$ on the eigenfuntions of  the sequence of Hamiltonians $H^{(j)}$, $j=0,1,2,\ldots$ In case that the selection of parameters $\epsilon_j$, $j=1,2,\ldots$ is appropriate and the transformation functions yield square integrable functions $\psi_{\epsilon_j}^{(j)}$, each supersymmetric transformation deforms the potential $V^{(j-1)}$ in such a way that the new one $V^{(j)}$ hosts exactly one additional eigenstate associated to the eigenvalue $\varepsilon_0^{(j)} = \epsilon_j$.}
    \label{fig:kSUSYQM}
\end{figure}

 After some calculations we obtain 
\begin{equation}\label{ec29.7}
     W_{2}({\rm x},\epsilon_1,\epsilon_{2})=-W_{1}({\rm x},\epsilon_{1})-\frac{2(\epsilon_{1}-\epsilon_{2})}{W_{1}({\rm x},\epsilon_{1})-W_{1}({\rm x},\epsilon_{2})}.
 \end{equation}
\begin{equation}\label{ec29.8}
    V^{(2)}({\rm x},\epsilon_1,\epsilon_2)=V^{(0)}(x)+2\partial_{\rm x}\left(\frac{\epsilon_{1}-\epsilon_{2}}{W_{1}({\rm x},\epsilon_{1})-W_{1}({\rm x},\epsilon_{2})}\right).
\end{equation}
with $W_1({\rm x},\epsilon_j) = \partial_{\rm x} \ln u^{(0)}({\rm x},\epsilon_j)$. Furthermore, the factorized forms of the Hamiltonians can be readily written
\begin{equation}\label{ec29.10}
    H^{(1)}(\epsilon_1)=L_{2}^{-}(\epsilon_1,\epsilon_2) L_{2}^{+}(\epsilon_1,\epsilon_2)+\epsilon_{2}, \qquad H^{(2)}(\epsilon_1,\epsilon_2)=L_{2}^{+}(\epsilon_1,\epsilon_2) L_{2}^{-}(\epsilon_1,\epsilon_2)+\epsilon_{2},
\end{equation}
where $L_{2}^{-}(\epsilon_1,\epsilon_2)=(L_{2}^{+}(\epsilon_1,\epsilon_2))^{\dagger}$, along with their eigenfunctions and eigenvalues
\begin{equation}
\label{2nd}
\begin{array}{ll}
  \displaystyle \psi_{\epsilon_{1}}^{(2)}({\rm x})=\frac{L_{2}^{+}(\epsilon_1,\epsilon_2)\,\psi_{\epsilon_{1}}^{(1)}({\rm x})}{\sqrt{\epsilon_{1}-\epsilon_2}},    &    \varepsilon_{1}^{(2)}=\varepsilon_{0}^{(1)}=\epsilon_{1}\\[2ex]
  \displaystyle \psi_{n+2}^{(2)}({\rm x}) = \frac{L_{2}^{+}(\epsilon_1,\epsilon_2) \, \psi_{n+1}^{(1)}({\rm x})}{\sqrt{\varepsilon_{n+1}^{(1)} - \epsilon_{2}}}= \frac{L_{2}^{+}(\epsilon_1,\epsilon_2) \,L_{1}^{+}(\epsilon_1) \, \psi_{n}^{(0)}({\rm x})}{\sqrt{\left(\varepsilon_{n}^{(0)} - \epsilon_{2}\right)\left(\varepsilon_{n}^{(0)} - \epsilon_{1}\right)}},    &   \varepsilon_{n+2}^{(2)}=\varepsilon_{n+1}^{(1)}=\varepsilon_{n}^{(0)}, \quad n = 0, 1, 2, \dots
\end{array}
\end{equation}
Moreover, the function
\begin{equation}\label{ec29.12}
\psi_{\epsilon_{2}}^{(2)}({\rm x}) = \mathcal{C}_2 \exp\left(-\int^{\rm x} W_{2}(t,\epsilon_1,\epsilon_{2}) \, \mathrm{d}t\right) = \frac{\mathcal{C}_2}{u^{(1)}({\rm x},\epsilon_1,\epsilon_2)},
\end{equation}
with $\mathcal{C}_2$ a (normalization) constant, must be considered among the solutions of the eigenvalue equation $H^{(2)}$ as it is a solution of the corresponding eigenvalue equation associated to the parameter $\varepsilon^{(2)}_0=\epsilon_2$. Notice that $\psi_{\epsilon_{2}}^{(2)}$ satisfies the equation $L_2^-(\epsilon_1,\epsilon_2) \psi_{\epsilon_2}^{(2)} = 0$, so that it is orthogonal to all eigenfunctions of the form \eqref{2nd}. If this function also fulfills the square integrability condition, then $\mathrm{Sp}\left(H^{(2)}\right) = \left\{\varepsilon_0^{(2)} = \epsilon_2, \varepsilon_1^{(2)} = \varepsilon_0^{(1)} = \epsilon_1, \varepsilon_{n+2}^{(2)} = \varepsilon_{n+1}^{(1)} = \varepsilon_n^{(0)},~n=0,1,2,\ldots\right\}$ contains exactly one and two additional elements to $\mathrm{Sp}\left(H^{(1)}\right)$ and $\mathrm{Sp}\left(H^{(0)}\right)$, respectively. In this case, $\mathcal{S}^{(2)} = \left\{\psi_{\epsilon_2}^{(2)},\psi_{\epsilon_1}^{(2)}, \psi_{n+2}^{(2)},~n=0,1,2,\ldots\right\}$ (see Figure \ref{fig:kSUSYQM}).
\subsubsection{\emph{k}-th order supersymmetric transformation}

The above iterative process can be developed up to the $k$-th order. Assuming the existence of $k$ linearly independent solutions $u^{(0)}(\epsilon_j)$, $j = 1,2,\ldots,k$ of the departure eigenvalue equation in \eqref{EigenEqH0} with $\varepsilon = \epsilon_j$ and $\epsilon_{j+1}<\epsilon_{j}$, we may construct the superpotentials $W_{1}(\epsilon_{j})$, such that $W_{1}({\rm x},\epsilon_{j})= \partial_{{\rm x}} \ln u^{(0)}({\rm x}, \epsilon_j)$. Hence, we obtain a sequence of Hamiltonians $H^{(j)}$ of the form \eqref{16} with potentials $V^{(j)}$ given by \cite{Mie00}
\begin{equation}\label{ec29.13}
V^{(j)}({\rm x},\epsilon_j)=V^{(j-1)}({\rm x},\epsilon_{j-1})-\partial^2_{\rm x} \ln u^{(0)}({\rm x},\epsilon_{j})=V^{(0)}({\rm x})-\sum_{s=1}^{j}\partial^2_{\rm x} \ln u^{(0)}_{s}({\rm x},\epsilon_{s}).
\end{equation}
Note that, although $H^{(j)}$, $V^{(j)}$ and $L_j^\pm$ depend on the whole set of parameters $\left\{\epsilon_1, \epsilon_2, \ldots, \epsilon_j\right\}$, in the previous expression it is only explicit the dependence of the last constant $\epsilon_j$. This notation will be used, from now on, for the sake of simplicity. Also, the intertwining operators $L_{j}^+$ can be always written in the form \eqref{ec29.1} where superpotentials $W_{j}({\rm x},\epsilon_{j})$ are constructed recursively via a finite difference formula similar to that in \eqref{ec29.7}:
 \begin{equation}\label{ec29.14}
     W_{j}({\rm x},\epsilon_{j})=-W_{j-1}({\rm x},\epsilon_{j-1})-\frac{2(\epsilon_{j-1}-\epsilon_{j})}{W_{j-1}({\rm x},\epsilon_{j-1})-W_{j-1}({\rm x},\epsilon_{j})}, \quad j=2,3,\dots,k.
 \end{equation}
 The Hamiltonians $H^{(j)}(\epsilon_j)$, $j = 0,1,2,\ldots,k$, where $H^{(0)}(\epsilon_0)= H^{(0)}$, are intertwined to their consecutive super-partners through the intertwining operators $L_j^+(\epsilon_j)$ in the form 
\begin{equation}\label{ec29.16}
    H^{(j)}(\epsilon_j)L_{j}^{+}(\epsilon_j)=L_{j}^{+}(\epsilon_j)H^{(j-1)}(\epsilon_j), \quad j=1,2,\dots,k,
\end{equation}
leading to the sequence of factorized deformed Hamiltonians  
\begin{equation}\label{ec29.17}
   H^{(j-1)}(\epsilon_{j-1})=L_{j}^{-}(\epsilon_{j})L_{j}^{+}(\epsilon_{j})+\epsilon_{j}, \qquad H^{(j)}(\epsilon_j) =L_{j}^{+}(\epsilon_j)L_{j}^{-}(\epsilon_j)+\epsilon_{j},  \qquad j=1,2,\dots,k.
\end{equation}
 Clearly, $\mathrm{Sp}\left(H^{(j)}\right) = \left\{\varepsilon_s^{(j)} =\epsilon_{j-s}, \quad s=0,1,\dots,j-1, \quad \varepsilon^{(j)}_{n+j} = \varepsilon_{n+j-1}^{(j-1)} = \cdots = \varepsilon_n^{(0)}, \quad n=0,1,2,\ldots\right\}$, $j=1,2,\ldots, k, $ and the corresponding set $\mathcal{S}^{(j)} = \left\{\psi_{\epsilon_j}^{(j)}, \psi_{\epsilon_{j-1}}^{(j)}, \ldots \psi_{\epsilon_1}^{(j)}, \psi_{n+j}^{(j)},~n=0,1,2\ldots\right\}$ of fundamental solutions of the eigenvalue problem are given by
 \begin{equation}
 \label{kth}
 \begin{array}{ll}
     \displaystyle \psi_0^{(j)}=\psi_{\epsilon_{j}}^{(j)}({\rm x})= \mathcal{C}_0\exp\left(-\int^{\rm x}W_{j}(t,\epsilon_{j})\,{\rm d}t\right),  &   \varepsilon_{0}^{(j)}=\epsilon_{j}, \\[3ex]
    \displaystyle \psi_s^{(j)}= \psi_{\epsilon_{j-s}}^{(j)}({\rm x})=\frac{L_{j}^{+}\,\psi_{\epsilon_{j-s}}^{(j-1)}({\rm x})}{\sqrt{\epsilon_{j-s}-\epsilon_{j}}}, &   \varepsilon_{s}^{(j)}=\epsilon_{j-s},  \quad  \quad s=1,\dots,j-1, \\[3ex]
      \displaystyle \psi_{n+j}^{(j)}({\rm x})=\frac{L_{j}^{+}\cdots L_{1}^{+}\,\psi_{n}^{(0)}({\rm x})}{\sqrt{(\varepsilon_{n}^{(0)}-\epsilon_{j})\cdots(\varepsilon_{n}^{(0)}-\epsilon_{1})}}, & \varepsilon_{n+j}^{(j)}=\varepsilon_{n}^{(0)}, \quad n=0,1,2,\dots, 
\end{array}
 \end{equation}
 with $j=1,2,\ldots,k$.

\section{Guided modes in a position-dependent refractive index}
\label{sec:headings}

Light is a manifestation of electromagnetic radiation composed of coupled, oscillating electric and magnetic fields that propagate through space and time. In many situations of practical interest, light may experience a beam-like behavior resembling an ideal, nonspreading, localized spatial distribution of electromagnetic energy. In this case, the propagation phenomena can be described within the \textit{paraxial approximation}, which assumes that light propagates along the longitudinal direction mostly as a plane wave, while remaining nearly confined in the transversal one \cite{Sie86,Sal91}. This approximation is valid for \emph{weakly guiding} materials having refractive profiles $n^{(0)}$ that exhibit slow variations in the transversal coordinates. Here, we assume that $n^{(0)}(\mathbf{q}) = n_\mathrm{B}  -\Delta n^{(0)}(\mathbf{q})$, with $\mathbf{q} = (x,y,z)$, $n_\mathrm{B}$ a constant reference refractive index, and $|\Delta  n^{(0)}(\mathbf{q})| \ll n_\mathrm{B}$, so that the weakly guiding condition is fulfilled. For a linearly polarized, harmonic beam-like perturbation propagating along the $z$-axis, the electric field is given by $\mathbf{E}^{(0)}(\mathbf{q};t) = \mathcal{E}\,\mathcal{U}^{(0)}(\mathbf{q}) \mathrm{e}^{i\left(k_0 n_\mathrm{B} z - \omega t \right)} \boldsymbol{\hat{e}}$, where $\mathcal{E}$ is a real constant, with electric field units, representing the field strength, $\boldsymbol{\hat{e}}$ stands for the (constant) polarization vector, $k_0$ for the wave number in free space, $\omega$ for the frequency, and the electric field enveloping amplitude $\mathcal{U}^{(0)}(\mathbf{q})$ satisfies the paraxial Helmholtz equation \cite{Cru15} 
\begin{equation}
-\frac{1}{2k_0^2 n_B}\nabla^2_{\rm \mathbf{r}} \, \mathcal{U}^{(0)}(\mathbf{q}) + \Delta n^{(0)}(\mathbf{q}) \, \mathcal{U}^{(0)}(\mathbf{q})=\frac{i}{k_0} \partial_\mathrm{z} \, \mathcal{U}^{(0)}(\mathbf{q}).
\label{ec.parax.sch}
\end{equation}
In this expression $\textbf{r}=(x,y)$, $\nabla_{\rm \mathbf{r}}^2= \partial^2_ x + \partial^2_y$ are, respectively, the position vector and the Laplacian in the transversal plane. In the paraxial approximation, it is assumed that the variations of $\mathcal{U}^{(0)}(\mathbf{q})= \mathcal{U}^{(0)}(\mathbf{r}, z)$ in the longitudinal coordinate $z$ are very small compared to those with respect to the transversal ones. Therefore, the dependence of $\mathcal{U}^{(0)}$ on the $(x, y)$ coordinates defines the transversal profile of the beam, while its dependence on $z$ modulates, very smoothly, the harmonic behavior of the fields. 

Note the resemblance of equation (\ref{ec.parax.sch}) to the time-dependent Schr\"odinger equation, a fact that has been widely used in diverse contexts \cite{Glo69,Sto81,Chu94,Dra02,Lon09,Cru15,Cru15a,Cru17,Lon18,Con19,Boc22,Cru23}. Furthermore, the square integrability condition, which in quantum mechanics allows the probabilistic interpretation of the wave function, in the optical context indicates that the energy carried by the beam is confined to a finite region of the transversal plane. In the wave optics approach, the irradiance or (normalized) optical power density is expressed as $I^{(0)}(\mathbf{q}) = \vert \mathcal{U}^{(0)}(\mathbf{q})\vert^2$ \cite{Sie86,Sal91}. Therefore, the square integrability condition
\begin{equation}
P^{(0)}=\int_{\mathbb{R}^2} I^{(0)}(\mathbf{q})\,{\rm d}x \, {\rm d}y = \int_{\mathbb{R}^2} \vert \mathcal{U}^{(0)}(\mathbf{q})\vert^2\,{\rm d}x \, {\rm d}y < \infty,
\label{power}
\end{equation}
ensures that the beam possesses a finite transversal optical power $P^{(0)}$ so that a finite amount of energy is required for its production in practice.

Let us assume that the refractive index depends only on the transversal variables, \emph{i.e.}, $\Delta n^{(0)}(\mathbf{q}) = \Delta n^{(0)}(\mathbf{r})$. Then, it is possible to find stationary solutions to (\ref{ec.parax.sch}) of the form 
\begin{equation}
\label{StatSol}
\mathcal{U}^{(0)}(\mathbf{q})= U^{(0)}(\mathbf{r}) \, {\rm e}^{-i k_0 \alpha \rm z},
\end{equation}
where $\alpha$ is a constant to be determined, and the amplitud $U^{(0)}$, defining the transversal distribution of optical power, satisfies the eigenvalue problem
\begin{equation}
\label{ec3}
H^{(0)} U^{(0)} = \alpha \, U^{(0)}.
\end{equation}
Here $H^{(0)}$ stands for the second order differential \emph{optical Hamiltonian} given by
\begin{equation}
\label{Ham0}
H^{(0)}= -\frac{1}{2k_0^2n_\mathrm{B}} \nabla_\mathbf{r}^2 + \Delta n^{(0)}(\mathbf{r}).
\end{equation}

The fields of the form (\ref{StatSol}) are equivalent to stationary states in quantum mechanics in the sense that their associated optical power is independent of the longitudinal coordinate. Check that the optical modes so constructed preserve the plane wave behavior as they propagate along the $z$-axis. The dimensionless constant $\alpha$ defines the speed at which each mode propagates and is called \emph{effective refractive index} of the mode. Moreover, the square integrability condition results in the discretization of $\alpha$ and, consequently, of the beam-like modes hosted by the weakly guiding optical medium described by the refractive index $n^{(0)}$. These beams are known as \emph{guided} or \emph{characteristic modes} of the material.    

\subsection{The eigenvalue problem for the cross-section amplitude}

Let us assume that the variations of $n^{(0)}$ in both transversal coordinates are independent of one another. This is true whenever the position-dependent term of the refractive index is written as
\begin{equation}\label{ec4}
n^{(0)}(\mathbf{r})= n_\mathrm{B} [1-f_x(x)-f_y(y)], 
\end{equation}
for some real-valued functions $f_{\xi}$ such that $\vert f_\xi(\xi) \vert \ll 1$, for $\xi \in \mathcal{D}_\xi = \mathrm{Dom}(f_\xi)$, $\xi = x,y$. The optical Hamiltonian $H^{(0)} = H^{(0)}_x + H^{(0)}_{y}$, where the one-dimensional Hamiltonians $H_\xi^{(0)}$ are given by 
\begin{equation}
\label{ec8}
H_{\xi}^{(0)}=-\frac{1}{2k_0^2n_\mathrm{B}}\partial^2_\xi +v^{(0)}_{\xi}(\xi), \qquad v^{(0)}_{\xi}(\xi)=n_{\mathrm B}f_{\xi}(\xi), \qquad \xi=x,y.
\end{equation} 
Thus the expression \eqref{ec3} separates into two one-dimensional eigenvalue problems, namely
\begin{equation}\label{ec6}
H^{(0)}_{x} X^{(0)}(x) = \alpha_{x} X^{(0)}(x), \qquad H^{(0)}_{y} Y^{(0)}(y) = \alpha_{y} Y^{(0)}(y),
\end{equation}
with $\alpha=\alpha_{x} + \alpha_{y}$ and  $U^{(0)}(\mathbf{r})= X^{(0)}(x)Y^{(0)}(y)$.

\subsection{Guided modes for an elliptic refractive index waveguide}

\begin{figure}[ht]
    \centering
    \includegraphics[width=0.55\linewidth]{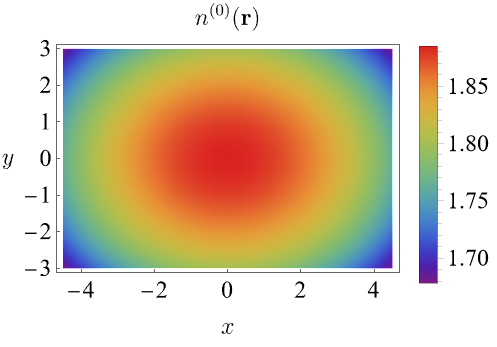}
\caption{Elliptic refractive index profile \eqref{ec9} for $n_{\mathrm B}=1.885$ with $\Omega_y = \frac54 \Omega_x$ and $\Omega_{x}= \frac{0.08}{w_0}$, $w_0$ being the width of the beam at the input (and focal) plane. Correspondingly, the transversal variables $x,y$ are measured in units of $w_0$.}
    \label{fig:n0_index}
\end{figure}

Refractive index distributions with quadratic dependences in the transversal coordinates have deserved a lot of interest in optics due to their confining, collimating and self-focusing properties \cite{Cru17,Cru19,Kot13,Pet16,Wu20}. Some other applications include the design of efficient modulation techniques in optical fiber communications \cite{Put21}. In this work, we will consider an optical medium with an elliptic refractive index profile with quadratic (but non necessarily symmetric) behavior in both transversal coordinates (see Figure \ref{fig:n0_index}):
\begin{equation}
n^{(0)}(\mathbf{r})= n_{\mathrm B}-\frac{n_{\mathrm B}}2\left(\Omega_{x}^2{x}^2+\Omega_{y}^2{y}^2\right), \qquad \Omega^2_\xi \xi^2 \ll 1 \qquad \xi=x,y
\label{ec9}
\end{equation}
where $\Omega_{x,y}$ are two positive constants characterizing the confining properties of the material in each transversal direction. Thus, the optical potential $v_\xi^{(0)}(\xi) = \frac12 n_\mathrm{B} \Omega_\xi \xi^2$ and the Hamiltonian (\ref{ec8}) becomes that of the quantum harmonic oscillator 
\begin{equation}\label{ec11}
H_{\xi}^{(0)}=- \frac{1}{2k_0^2\,n_\mathrm{B}} \partial^2_\xi + \frac12 n_\mathrm{B} \Omega^2_\xi \xi^2
\end{equation}

\begin{figure}[htbp]
     \centering
     \begin{subfigure}{0.32\linewidth}
         \centering
    \includegraphics[width=\linewidth]{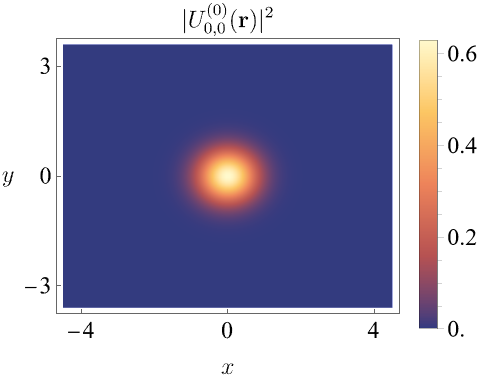}
        \caption{}
        \label{fig:4a}
     \end{subfigure}
     \hfill
     \begin{subfigure}{0.32\linewidth}
         \centering
    \includegraphics[width=\linewidth]{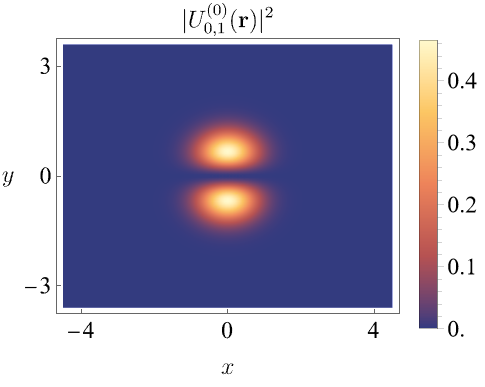}
        \caption{}
        \label{fig:4b}
     \end{subfigure}
     \hfill
     \begin{subfigure}{0.32\linewidth}
         \centering
        \includegraphics[width=\linewidth]{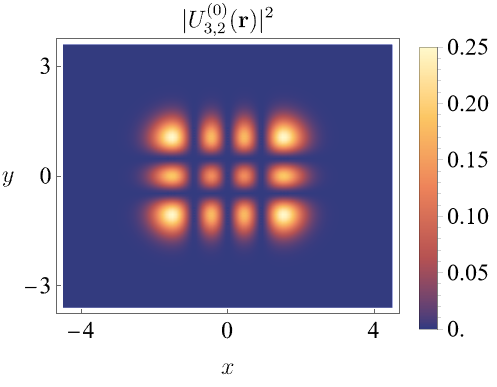}
        \caption{}
        \label{fig:4c}
     \end{subfigure}
    \caption{Transversal intensity distributions of the GHG modes with $n_\mathrm {B}=1.885$, $\Omega_{y}=\frac54\Omega_{x}$ and $\Omega_x = \frac{0.08}{w_0}$: (a) $\vert U_{0,0}^{(0)}(\mathbf{r})\vert^2$, (b) $\vert U_{0,1}^{(0)}(\mathbf{r})\vert^2$, (c) $\vert U_{3,2}(\mathbf{r})^{(0)}\vert^2$. In all cases, the transversal coordinates $x,y$ are measured in units of $w_0$.}
    \label{fig:fig4}
\end{figure}

It is well known that the parabolic profile supports diverse families of stationary and non-stationary localized beams \cite{Cru17,Cru19,Cru23}. Yet, we will fix attention on the particular family of guided (stationary) Hermite-Gaussian (GHG) modes. Each mode satisfies the square-integrable condition \eqref{power} so that they conform the set of fundamental (normalized) solutions to the 2D separable eigenvalue problem (\ref{ec6}) $\mathcal{S}^{(0)} = \left\{U^{(0)}_{n,m}:~n,m=0,1,2,\ldots\right\}$ with \cite{Cru17,Gre17,Cru19}
\begin{equation}
\label{GHG}
U_{n,m}^{(0)}(\mathbf{r})=X_{n}^{(0)}(x) Y_{m}^{(0)}(y)=\sqrt{\frac{k_0\,n_\mathrm{B}\,\sqrt{\Omega_{x}\Omega_{y}}}{2^{n+m} n! m! \pi}} {\rm e}^{-\frac{1}{2}k_0 \, n_\mathrm{B}(\Omega_{x}{x}^2+\Omega_{y}{y}^2)} \mathbf{H}_n\left(\sqrt{k_0\,n_{\mathrm B} \Omega_{x}}x\right) \mathbf{H}_m\left(\sqrt{k_0\,n_{\mathrm B} \Omega_{y}}y\right).
\end{equation}
Here $\mathbf{H}_n(\mathrm{z})$ stands for the Hermite polynomials of degree $n$. In turn, the associated set of eigenvalues conforms the spectrum ${\rm Sp}\left(H^{(0)}\right)$ of \emph{effective refractive indices} 
\begin{equation}
\label{sp}
{\rm Sp}\left(H^{0}\right) = \left\{\alpha_{n,m}=\frac{1}{k_0} \left[\Omega_x\left(n+\frac12\right)+\Omega_{y}\left(m+\frac12\right)\right], \quad n,m=0,1,2,\ldots\right\},
\end{equation}
one for each guided mode in $\mathcal{S}^{(0)}$. 
In  Figure \ref{fig:fig4} we present the transversal intensity distributions of GHG modes for different values of the parameters $(n,m)$ and the same values of $n_\mathrm{B}$, $\Omega_x$ and $\Omega_y$ of Figure \ref{fig:n0_index}. Check the rectangular patterns possessing $n$, $m$ nodes, respectively, in the $x$ and $y$ directions.

\section{Supersymmetric partners of the parabolic medium}
\label{SusyO}

In this section, we apply the supersymmetric formalism to the elliptic refractive index in order to obtain new refractive profiles sharing similar spectral properties. For this purpose, it will be convenient to introduce new dimensionless variables $\chi(\xi) =\sqrt{k_0 n_{\rm B} \Omega_\xi} \, \xi$, $\xi = x,y$. With this change of variable, the Hamiltonian $H_\xi^{(0)}$ in \eqref{ec11} transforms into
\begin{equation}\label{ec30}
H_\chi^{(0)}= -\frac12\partial^2_\chi +\frac12 \chi^2, 
\end{equation}
that has the form \eqref{16} with $V^{(0)}(\chi) = \frac12 \chi^2$.

As discussed in Section \ref{SusyQM}, the key ingredient in the supersymmetric transformation is the selection of the transformation function $u^{(0)}$ that, according to \eqref{seedfunction}, is a solution of the eigenvalue equation $H_\chi^{(0)} u^{(0)} = \epsilon_\xi u^{(0)}$, with $\alpha_\xi = \frac{\Omega_\xi}{k_0} \, \epsilon_\xi$.  The general solution to this equation for the potential $V^{(0)}(\chi)$, has the form
\begin{equation}
u^{(0)}(\chi,\epsilon_\xi)={\rm e}^{\chi^2/2}\left[A \,_{1}F_{1}\left(\frac{1+2\epsilon_{\xi}}{4}, \frac{1}{2}; -\chi^2\right) +B\,\chi\,_{1}F_{1}\left(\frac{3+2\epsilon_\xi}{4}, \frac{3}{2};-\chi^2\right)\right], \label{ec32}
\end{equation}
where $\!_1F_1(a,b;z)$ is the confluent hypergeometric function and $A,B$ are arbitrary constants. Of course, for the choices $\epsilon_x = n + \frac12$, $\epsilon_y = m+\frac12$, and the suitable values of $A,B$, we obtain the eigenamplitudes $X_n(x) = u^{(0)}(\chi(x),\epsilon_x)$ and $Y_m(y) = u^{(0)}(\chi(y),\epsilon_y)$ leading to the characteristic modes \eqref{GHG} of the elliptic profile.
Nonetheless, in order to avoid the appearing of singularities in the new potential $V^{(1)}(\chi,\epsilon_\xi) = V^{(0)}(\chi) - \partial^2_\chi\ln u^{(0)}(\chi,\epsilon_\xi)$, the constants $A$ and $B$ must be suited so that the transformation function does not vanish for all $\chi \in \mathbb{R}$ and $\epsilon_\xi < \frac12$. After a simple analysis of the asymptotic behavior of \eqref{ec32}, and setting $A=1$, we obtain \cite{Mie84,Jun98,Var23}:
\begin{equation} 
u^{(0)}(\chi,\epsilon_\xi) = {\rm e}^{\chi^2/2}\Bigg[\,_1F_1\left(\dfrac{1 + 2\epsilon_\xi}{4}, \dfrac{1}{2}; -\chi^2 \right) + \dfrac{2\Gamma\left(\frac{3 - 2\epsilon_\xi}{4}\right)}{\Gamma\left(\frac{1 - 2\epsilon_\xi}{4} \right)} \, \gamma _\xi\, \chi\, 
_1F_1\left( \dfrac{3 + 2\epsilon_\xi}{4}, \dfrac{3}{2}; -\chi^2 \right)\Bigg], \label{ec33}
\end{equation}
where $\vert\gamma_\xi\vert<1$.
\begin{figure}[ht]
     \centering
     \begin{subfigure}{0.4\linewidth}
         \centering
    \includegraphics[width=\linewidth]{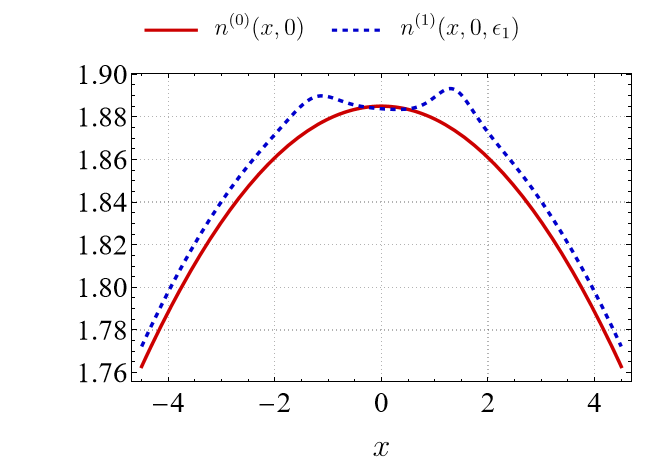}
        \caption{}
        \label{fig:5a}
     \end{subfigure}
     \hskip1cm
     \begin{subfigure}{0.4\linewidth}
         \centering
    \includegraphics[width=\linewidth]{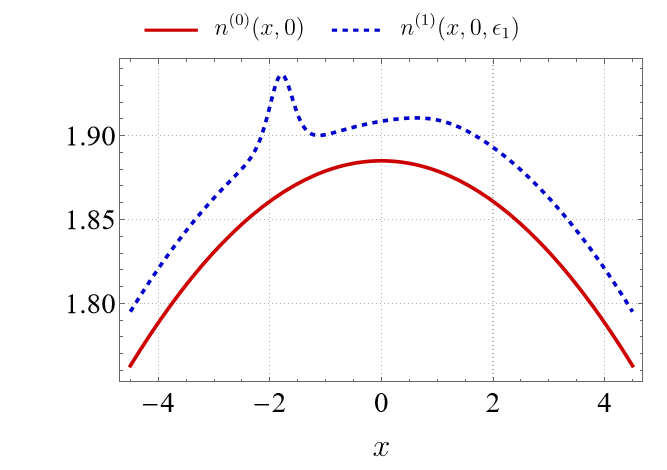}
        \caption{}
        \label{fig:5b}
     \end{subfigure}
    \caption{Transversal profile $n^{(0)}(\mathbf{r})$ (solid red line) and supersymmetric partners $n^{(1)}(\mathbf{r})$ (dotted blue line) at $y=0$ constructed with (a) $\epsilon_{x}=\frac14$, $\epsilon_y =-\frac14$, $\gamma_{x}=-\frac12$ and $\gamma_y = \frac12$, (b) $\epsilon_{x}=-\frac12$, $\epsilon_{y}=-\frac32$, $\gamma_{x}=0.9990$ and $\gamma_y =0$. In both cases, the transversal coordinate $x$ is measured in units of $w_0$.}
    \label{fig:potencial_socio}
\end{figure}

Thus, by choosing the set of parameters $\left\{\epsilon_\xi, \gamma_\xi, \xi= x,y\right\}$ and using \eqref{V1u} and \eqref{ec4}, we arrive at
$$
v^{(1)}(\xi,\epsilon_\xi) = V^{(1)}(\chi(\xi),\epsilon_\xi) = \frac12 n_\mathrm{B} \Omega_\xi \xi^2 - \frac1{k_0^2n_{\rm B}} \partial^2_\xi \ln u^{(0)}(\chi(\xi),\epsilon_\xi), 
$$
and
\begin{equation}\label{ec34}
n^{(1)}(\mathbf{r}) = n_{\rm B} - \frac12 n_{\rm B} \left(\Omega_x^2 x^2 + \Omega_y^2 y^2\right) + \frac1{k_0^2n_{\rm B}} \left(\partial^2_x \ln u^{(0)}(\sqrt{ k_0n_{\rm B} \Omega_x} \, x) + \partial^2_y \ln u^{(0)}(\sqrt{ k_0n_{\rm B} \Omega_y} \, y)\right).
\end{equation}
\begin{figure}[htbp]
     \centering
     \begin{subfigure}{0.4\linewidth}
         \centering
    \includegraphics[width=\linewidth]{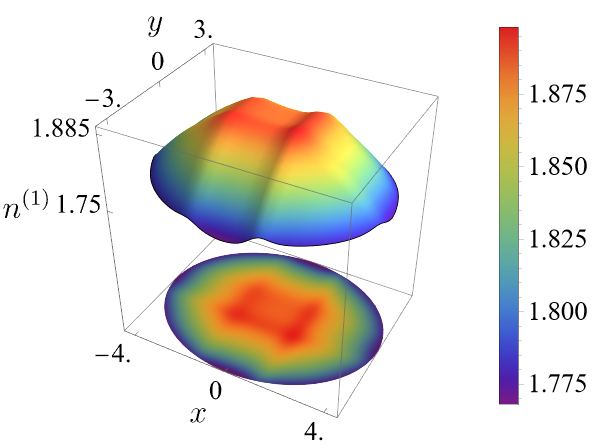}
        \caption{}
        \label{fig:6a}
     \end{subfigure}
     \hskip1cm
     \begin{subfigure}{0.4\linewidth}
         \centering
    \includegraphics[width=\linewidth]{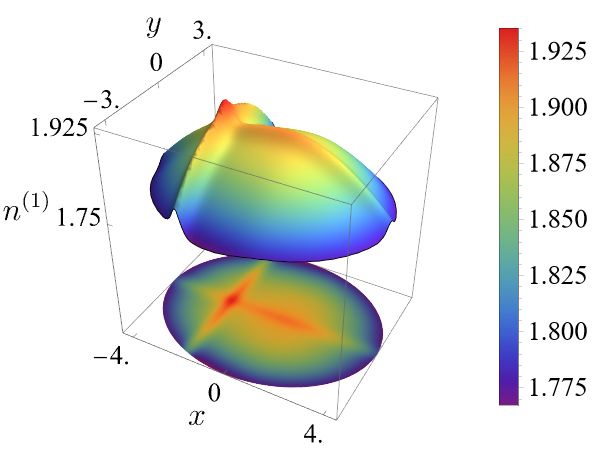}
        \caption{}
        \label{fig:6b}
     \end{subfigure}
    \caption{3D profile of the supersymmetric partners $n^{(1)}(\mathbf{r})$ constructed with (a) $\epsilon_{x}=\frac14$, $\epsilon_y =-\frac14$, $\gamma_{x}=-\frac12$ and $\gamma_y = \frac12$, (b) $\epsilon_{x}=-\frac12$, $\epsilon_{y}=-\frac32$, $\gamma_{x}=0.999$ and $\gamma_y =0$. In both cases, the transversal coordinates $x,y$ are measured in units of $w_0$.}
    \label{fig:potencial_socio1}
\end{figure}

In Figures \ref{fig:potencial_socio} and \ref{fig:potencial_socio1} we present some views of supersymmetric partners of the elliptic refractive index for different choices of the set $\left\{\epsilon_\xi, \gamma_\xi, \xi= x,y\right\}$. The multiparametric expression \eqref{ec34} allows the manipulation of the so-constructed profile in diverse ways. Check that the supersymmetric transformation introduces a peaked deformation on the departure profile. This deformation is consistent with the fact that we are inserting new localized modes out of those already hosted by $n^{(0)}$. Thus, this peak is more pronounced as $\vert \epsilon_\xi\vert$ becomes larger. The spatial position and sharpness of this peak are controlled by the values of $\gamma_\xi$. The peak becomes sharpest and goes far from the origin as the value of $\vert \gamma_\xi \vert$ approaches 1. Meanwhile, the position of the new spectral value $\alpha_{00}$ is defined by the choice of the parameters $\epsilon_\xi$ such that $\epsilon_\xi \leq \frac12$. 

\begin{figure}[htbp]
     \centering
     \begin{subfigure}{0.4\linewidth}
         \centering
    \includegraphics[width=\linewidth]{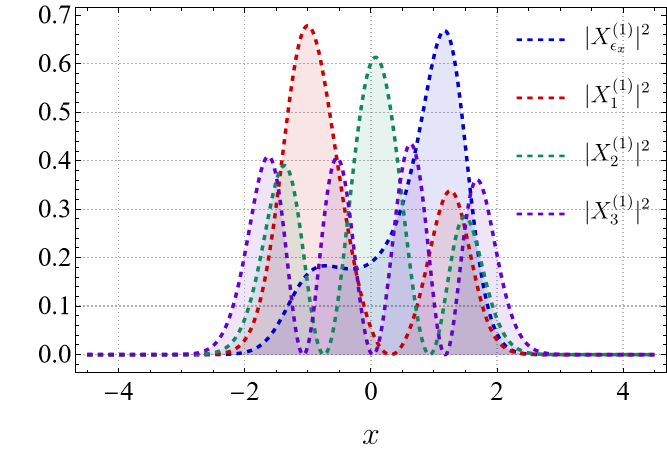}
        \caption{}
        \label{fig:7a}
     \end{subfigure}
     \hskip1cm
     \begin{subfigure}{0.4\linewidth}
         \centering
    \includegraphics[width=\linewidth]{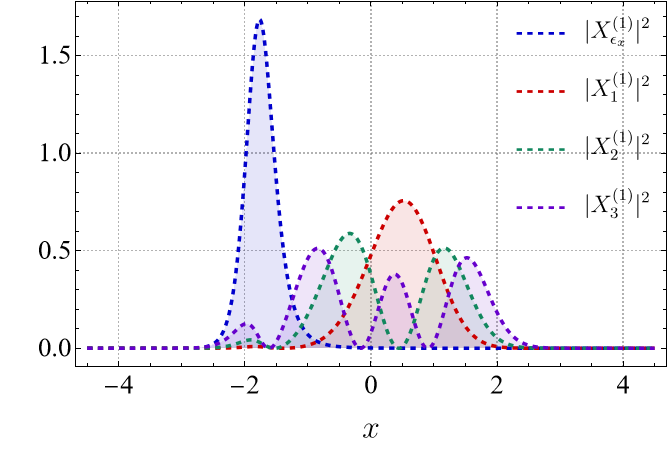}
        \caption{}
        \label{fig:7b}
     \end{subfigure}
    \caption{Square modulus of the amplitudes $X_{\epsilon_{x}}^{(1)}$ and $X_{n+1}^{(1)}$, $n=0,1,2$ constructed with (a) $\epsilon_{x}=1/4$, $\gamma_{x}=-1/2$ and (b) $\epsilon_{x}=-1/2$, $\gamma_{x}=0.999$. In both cases, the transversal coordinate $x$ is measured in units of $w_0$.}
    \label{fig:states_socio}
\end{figure}

\begin{figure}[htbp]
     \centering
     \begin{subfigure}{0.32\linewidth}
         \centering
    \includegraphics[width=\linewidth]{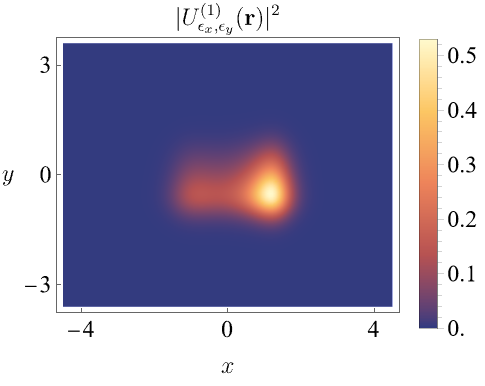}
        \caption{}
        \label{fig:8a}
     \end{subfigure}
     \hfill
     \begin{subfigure}{0.32\linewidth}
         \centering
    \includegraphics[width=\linewidth]{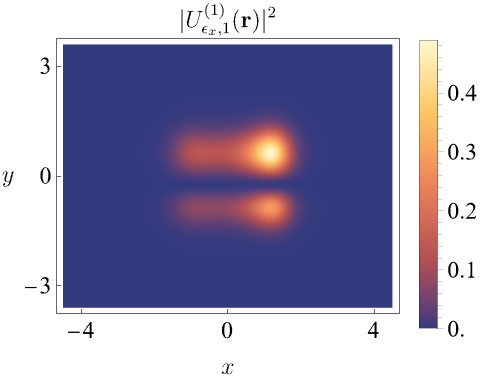}
        \caption{}
        \label{fig:8b}
     \end{subfigure}
     \hfill
     \begin{subfigure}{0.32\linewidth}
         \centering
        \includegraphics[width=\linewidth]{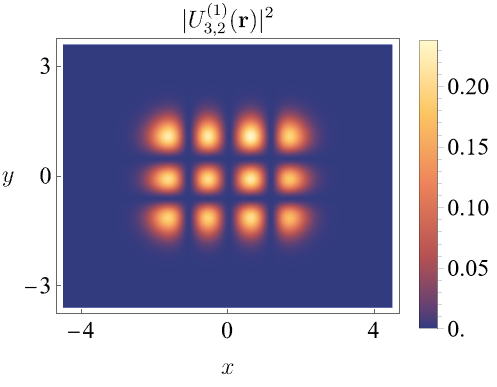}
        \caption{}
        \label{fig:8c}
     \end{subfigure}
    \caption{Transversal intensity distributions of the GHG modes with $n_\mathrm {B}=1.885$, $\Omega_{y}=\frac54\Omega_{x}$, $\Omega_x = \frac{0.8}{w0}$, $\epsilon_x = \frac14$, $\epsilon_y = -\frac14$, $\gamma_x = -\frac12$, $\gamma_y = \frac12$: (a) $\vert U_{\epsilon_x,\epsilon_y}^{(1)}(\mathbf{r})\vert^2$, (b) $\vert U_{\epsilon_x,1}^{(1)}(\mathbf{r})\vert^2$, (c) $\vert U_{3,2}^{(1)}(\mathbf{r})\vert^2$. In all cases, the transversal coordinates $x,y$ are measured in units of $w_0$.}
    \label{fig:fig8}
\end{figure}

\begin{figure}[htbp]
     \centering
     \begin{subfigure}{0.32\linewidth}
         \centering
    \includegraphics[width=\linewidth]{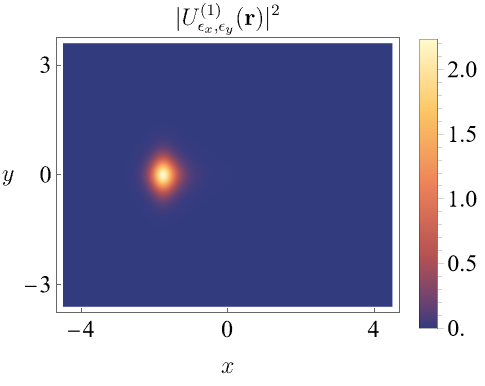}
        \caption{}
        \label{fig:9a}
     \end{subfigure}
     \hfill
     \begin{subfigure}{0.32\linewidth}
         \centering
    \includegraphics[width=\linewidth]{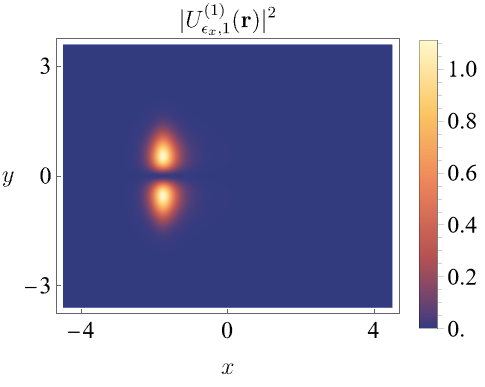}
        \caption{}
        \label{fig:9b}
     \end{subfigure}
     \hfill
     \begin{subfigure}{0.32\linewidth}
         \centering
        \includegraphics[width=\linewidth]{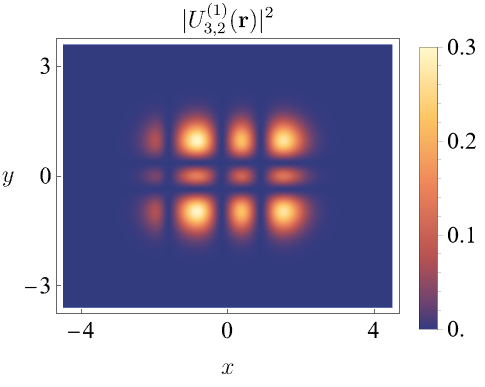}
        \caption{}
        \label{fig:9c}
     \end{subfigure}
    \caption{Transversal intensity distributions of the GHG modes with $n_\mathrm {B}=1.885$, $\Omega_{y}=\frac54\Omega_{x}$, $\Omega_x = \frac{0.8}{w_0}$, $\epsilon_x = -\frac12$, $\epsilon_y = -\frac32$. $\gamma_x = 0$ and $\gamma_y = 0.999$ (a) $\vert U_{\epsilon_x,\epsilon_y}^{(1)}(\mathbf{r})\vert^2$, (b) $\vert U_{\epsilon_x,1}^{(1)}(\mathbf{r})\vert^2$, (c) $\vert U_{3,2}^{(1)}(\mathbf{r})\vert^2$. In all cases, the transversal coordinates $x,y$ are measured in units of $w_0$.}
    \label{fig:fig9}
\end{figure}

In turn, the expressions for the amplitude $U^{(1)}_{n+1,m+1}(\mathbf{r})$, $n,m=0,1,2,\ldots$ of the characteristic modes are determined through equation \eqref{ec28}, where $X_{n+1}^{(1)}(x) = \psi_{n+1}^{(1)}(\chi(x))$ and $Y_{m+1}^{(1)}(y) = \psi_{m+1}^{(1)}(\chi(y))$. Additionally, the form of the missing modes $U^{(1)}_{\epsilon_x,\epsilon_y}$, $U^{(1)}_{\epsilon_x,m+1}$, and $U^{(1)}_{n+1,\epsilon_y}$ can be readily written from \eqref{ec29} once the parameters $\left\{\epsilon_\xi, \gamma_\xi, \xi= x,y\right\}$ are fixed (see Figures \ref{fig:states_socio}-\ref{fig:fig9}). 

\begin{figure}[htbp]
     \centering
     \begin{subfigure}{0.40\linewidth}
         \centering
    \includegraphics[width=\linewidth]{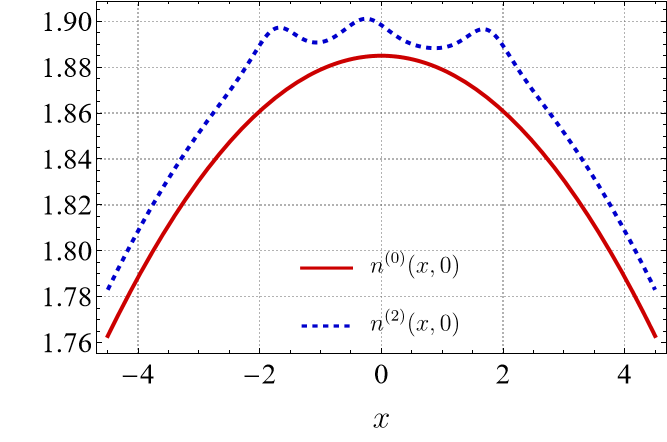}
        \caption{}
        \label{fig:10a}
     \end{subfigure}
     \hskip1cm
     \begin{subfigure}{0.40\linewidth}
         \centering
    \includegraphics[width=\linewidth]{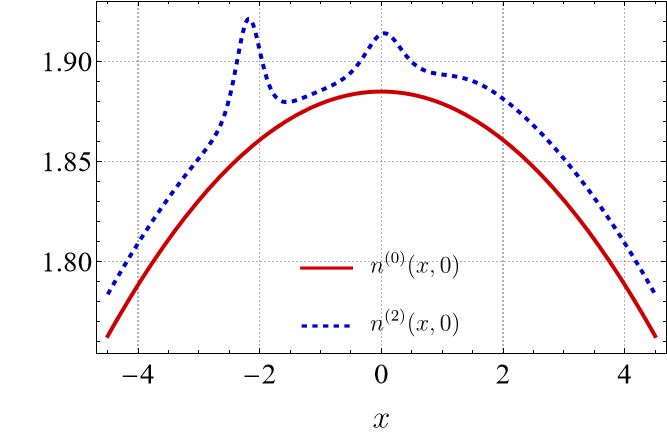}
        \caption{}
        \label{fig:10b}
     \end{subfigure}
    \caption{Transversal projection of $n^{(0)}(\mathbf{r})$ (solid red line) and its second order supersymmetric $n^{(2)}(\mathbf{r})$ (dotted blue line) at $y=0$ constructed with (a) $\epsilon_{1x}= \frac14$, $\epsilon_{2x}= \epsilon_{1y} = -\frac12$, $\epsilon_{2y} = -\frac32$, $\gamma_{1x} =\gamma_{1y}=0$, $\gamma_{2x} = 2$ and $\gamma_{2y} = -1$ (b) $\epsilon_{1x}=\epsilon_{1y}=-\frac12$, $\epsilon_{2x} = \epsilon_{2y} = -\frac32$, $\gamma_{1x} = \gamma_{1y}=0.999$,  $\gamma_{2x} = \gamma_{2y}=2$. In both cases the transversal coordinate $x$ is measured in units of $w_0$.}
    \label{fig:potencial_socio_2}
\end{figure}

\begin{figure}[htbp]
     \centering
     \begin{subfigure}{0.4\linewidth}
         \centering
    \includegraphics[width=\linewidth]{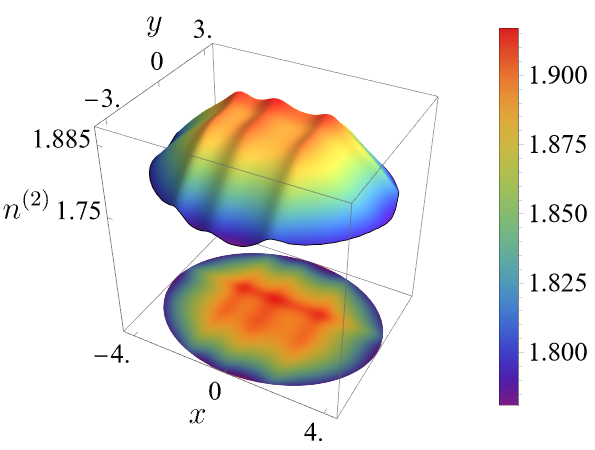}
        \caption{}
        \label{fig:11a}
     \end{subfigure}
     \hskip1cm
     \begin{subfigure}{0.4\linewidth}
         \centering
    \includegraphics[width=\linewidth]{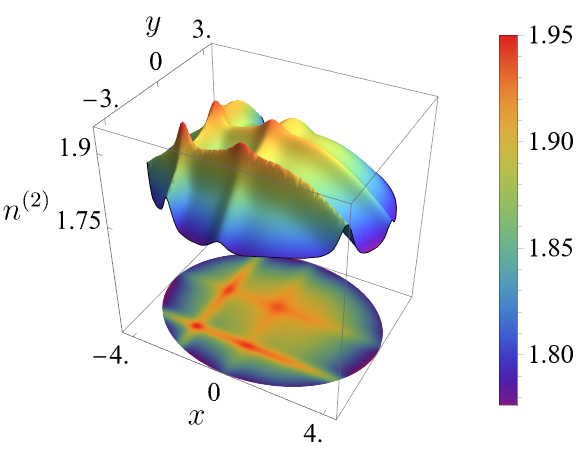}
        \caption{}
        \label{fig:11b}
     \end{subfigure}
    \caption{3D profile of the second order supersymmetric partner $n^{(2)}(\mathbf{r})$ constructed with (a) $\epsilon_{1x}= \frac14$, $\epsilon_{2x}= \epsilon_{y1} = -\frac12$, $\epsilon_{2y} = -\frac32$, $\gamma_{1x} =\gamma_{1y}=0$, $\gamma_{2x} = 2$ and $\gamma_{2y} = -1$ (b) $\epsilon_{1x}=\epsilon_{1y}=-\frac12$, $\epsilon_{2x} = \epsilon_{2y} = -\frac32$, $\gamma_{1x} = \gamma_{1y}=0.999$,  $\gamma_{2x} = \gamma_{2y}=2$. In both cases, the transversal coordinates $x,y$ are measured in units of $w_0$.}
    \label{fig:potencial_socio2}
\end{figure}

The second order supersymmetric partner of $v^{(2)}(\xi,\epsilon_{\xi})$ can be obtained using \eqref{ec29.6}. Let $\epsilon_\xi = \epsilon_{1\xi}$. Thus
\begin{equation}
\label{susy2eq}
v^{(2)}(\xi,\epsilon_{1\xi},\epsilon_{2\xi}) = \frac12 n_{\rm B} \Omega_\xi^2 \xi^2 -  \frac2{k_0n_{\rm B}} \partial^2_\xi \ln \mathcal{W}\left(u^{(0)}(\epsilon_{1\xi}),u^{(0)}(\epsilon_{2\xi})\right),
\end{equation}
where $u^{(0)}(\epsilon_{1\xi}),u^{(0)}(\epsilon_{2\xi})$, $\epsilon_{1\xi} > \epsilon_{2\xi}$, are two linearly independent solutions of the form \eqref{ec33}. The second order supersymmetric profile can be readily computed to yield (see Figures \ref{fig:potencial_socio_2} and \ref{fig:potencial_socio2}):
\begin{align} 
n^{(2)}(\mathbf{r},\epsilon_{1x},\epsilon_{2x},\epsilon_{1y},\epsilon_{2y}) & =n_B  - \frac12 n_{\rm B} \left(\Omega_x^2 x^2 + \Omega_y^2 y^2\right) + \frac2{k_0^2n_{\rm B}} \partial^2_x \ln \left[\mathcal{W}\left(u^{(0)}(x,\epsilon_{1x}),u^{(0)}(x,\epsilon_{2x})\right)\right] \nonumber \\ 
&\quad+ \frac2{k_0^2n_{\rm B}} \partial^2_y \ln \left[\mathcal{W}\left(u^{(0)}(y,\epsilon_{1y}),u^{(0)}(y,\epsilon_{2y})\right)\right]. \label{ec43}
\end{align}
\begin{figure}[htbp]
     \centering
     \begin{subfigure}{0.4\linewidth}
         \centering
    \includegraphics[width=\linewidth]{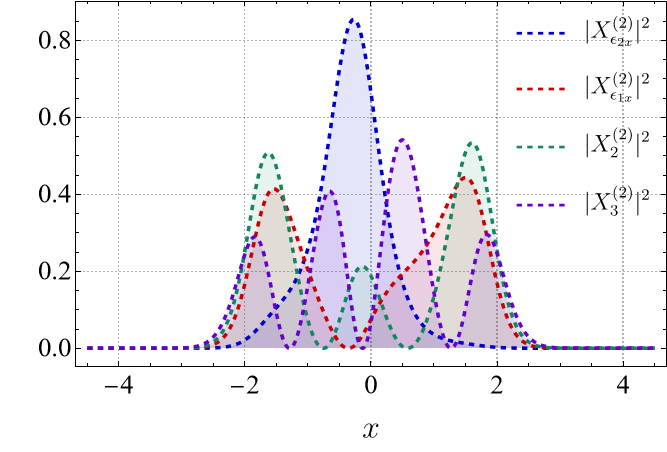}
        \caption{}
        \label{fig:12a}
     \end{subfigure}
     \hskip1cm
     \begin{subfigure}{0.4\linewidth}
         \centering
    \includegraphics[width=\linewidth]{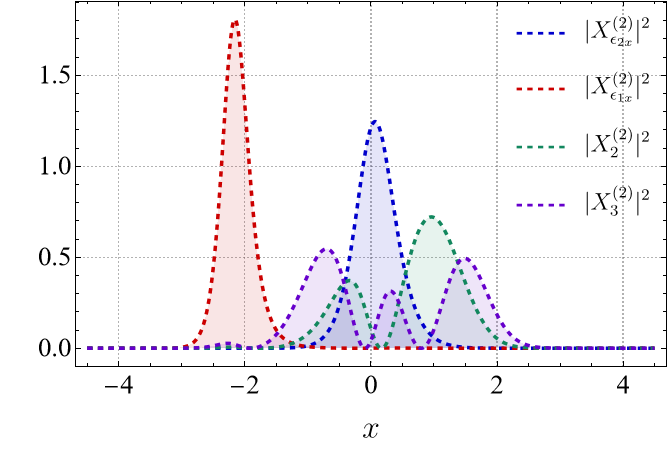}
        \caption{}
        \label{fig:12b}
     \end{subfigure}
    \caption{Square modulus of the enveloping functions $X^{(2)}_{\epsilon_{2x}}$, $X^{(2)}_{\epsilon_{1x}}$ and $X^{(2)}_{n+2}$, $n=0,1$ constructed with (a) $\epsilon_{1x}=\frac14$,  $\epsilon_{2x}= -\frac12$, $\gamma_{1x}=0$ and (b) $\epsilon_{1x}= -\frac12, \epsilon_{2x}= -\frac32$, $\gamma_{1x}=0.999$. In both cases, $\gamma_{2,x}=2$ and the transversal coordinates $x,y$ are measured in units of $w_0$.}
    \label{fig:states_socio_2}
\end{figure}

\begin{figure}[h]
     \centering
     \begin{subfigure}{0.32\linewidth}
         \centering
    \includegraphics[width=\linewidth]{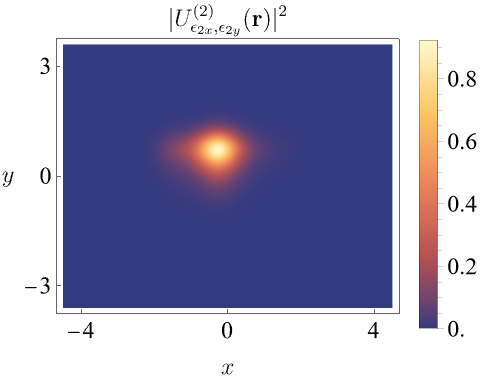}
        \caption{}
        \label{fig:13a}
     \end{subfigure}
     \hfill
     \begin{subfigure}{0.32\linewidth}
         \centering
    \includegraphics[width=\linewidth]{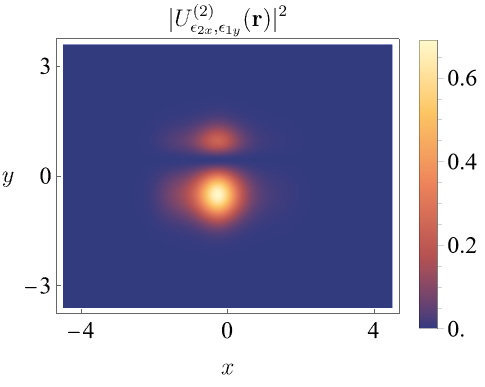}
        \caption{}
        \label{fig:13b}
     \end{subfigure}
     \begin{subfigure}{0.32\linewidth}
         \centering
        \includegraphics[width=\linewidth]{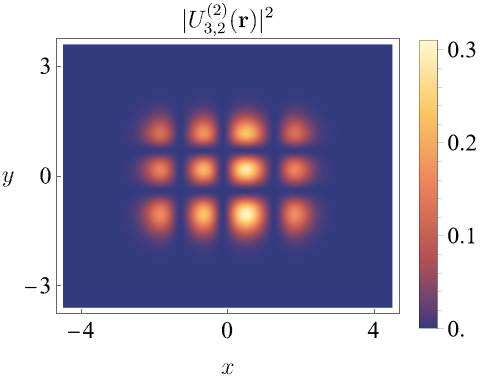}
        \caption{}
        \label{fig:13c}
     \end{subfigure}
     \hfill
    \caption{Transversal intensity distributions of the GHG modes with $n_\mathrm {B}=1.885$, $\Omega_{y}=\frac54\Omega_{x}$, $\Omega_x = \frac{0.8}{w_0}$, $\epsilon_{1x} = \frac14$, $\epsilon_{2x} = \epsilon_{1y} = -\frac12$, $\epsilon_{2y} = -\frac32$, $\gamma_{1x} = \gamma_{1y} = 0$, $\gamma_{2x} = 2$ and $\gamma_{2y} = -1.2$ (a) $\vert U_{\epsilon_{2x},\epsilon_{2y}}^{(2)}(\mathbf{r})\vert^2$, (b) $\vert U_{\epsilon_{2x},\epsilon_{1y}}^{(2)}(\mathbf{r})\vert^2$, (c) $\vert U_{3,2}^{(2)}(\mathbf{r})\vert^2$. In all cases, the transversal coordinates $x,y$ are measured in units of $w_0$.}
    \label{fig:Modos4_SUSY}
\end{figure}

\begin{figure}[htbp]
     \centering
     \begin{subfigure}{0.32\linewidth}
         \centering
    \includegraphics[width=\linewidth]{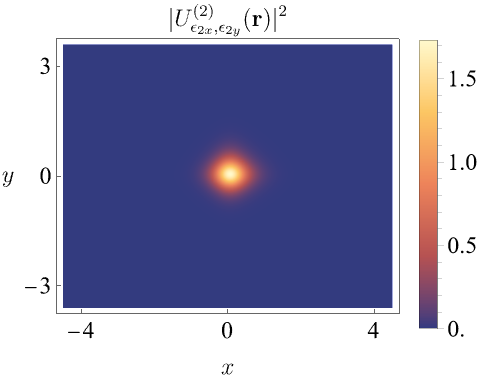}
        \caption{}
        \label{fig:14a}
     \end{subfigure}
     \hfill
     \begin{subfigure}{0.32\linewidth}
         \centering
    \includegraphics[width=\linewidth]{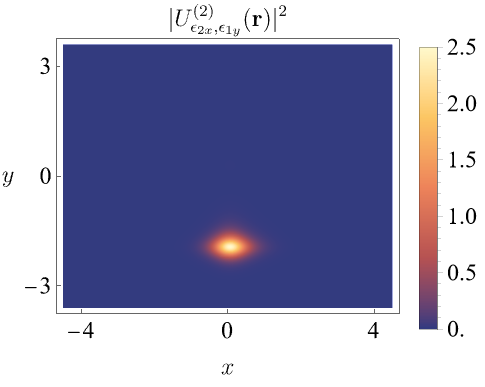}
        \caption{}
        \label{fig:14b}
     \end{subfigure}
     \begin{subfigure}{0.32\linewidth}
         \centering
        \includegraphics[width=\linewidth]{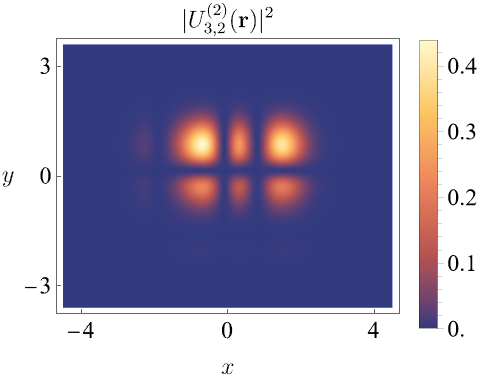}
        \caption{}
        \label{fig:14c}
     \end{subfigure}
     \hfill
    \caption{Transversal intensity distributions of the GHG modes with $n_\mathrm {B}=1.885$, $\Omega_{y}=\frac54\Omega_{x}$, $\Omega_x = \frac{0.8}{w_0}$, $\epsilon_{1x} = \epsilon_{1y} = -\frac12$, $\epsilon_{2x} = \epsilon_{2y} = -\frac32$, $\gamma_{1x} = \gamma_{1y} = 0.999$ and $\gamma_{2x} = \gamma_{2y} = 2$ (a) $\vert U_{\epsilon_{2x},\epsilon_{2y}}^{(2)}(\mathbf{r})\vert^2$, (b) $\vert U_{\epsilon_{2x},\epsilon_{1y}}^{(2)}(\mathbf{r})\vert^2$, (c) $\vert U_{3,2}^{(2)}(\mathbf{r})\vert^2$. In all cases, the transversal coordinates $x,y$ are measured in units of $w_0$.}
    \label{fig:Modos5_SUSY}
\end{figure}

Correspondingly, the field amplitudes $U^{(2)}(\mathbf{r})$ are determine through expression \eqref{2nd}. In Figure \ref{fig:states_socio_2} we present the transversal profile of the $x$-projection of the enveloping functions $X^{(2)}_{\epsilon_{2x}}$, $X^{(2)}_{\epsilon_{1x}}$, $X^{(2)}_2$ and $X^{(2)}_3$ for the same values of the parameters used in Figure \ref{fig:potencial_socio2}. In turn, Figures \ref{fig:Modos4_SUSY} and \ref{fig:Modos5_SUSY} show the transversal distributions $U^{(2)}(\mathbf{r})$ for different values of the parameters $\{\epsilon_{\xi},\gamma_{\xi},\xi=x,y\}$.

The supersymmetric transformations, enables us to obtain solutions of the sequence of stationary paraxial Helmholtz equation
\begin{equation}\label{ec46}
-\frac{1}{2k_0^2 n_0}\nabla^2_{\rm \mathbf{r}} U^{(j)}(\mathbf{r}) + \left(n_0-n^{(j)}(\mathbf{r})\right) U^{(j)}(\mathbf{r})=\alpha\, U^{(j)}(\mathbf{r}), \quad j=1,2,\ldots
\end{equation}
where each profile $n^{(j)}(\mathbf{r})$, $j=1,2,\ldots$, hosts one more guided mode than its predecessor  $n^{(j-1)}(\mathbf{r})$. For instance, in Figures \ref{fig:potencial_socio1} and \ref{fig:potencial_socio2}, we plot refractive index profiles $n^{(1)}(\mathbf{r})$ and $n^{(2)}(\mathbf{r})$ constructed with the partner potentials $V_{x,y}^{(1)}$ and $V_{x,y}^{(2)}$, respectively, that exhibit clear differences with respect to that in equation \eqref{ec9} (see Figure \ref{fig:n0_index}). In general, the degree of asymmetry is greater in these last profiles. Also, it is worth remarking that, in order to avoid singularities in $V_{x,y}^{(2)}$, the parameters $\gamma_{2,x}$ and $\gamma_{2,y}$ must fulfill $\gamma_{2,x},\gamma_{2,y}\in\mathbb{R}-\{(-1,1)\}$.

Figures \ref{fig:fig8}, and \ref{fig:fig9}, show some guided modes hosted by the profiles $n^{(1)}(\mathbf{r})$ for some choices of the parameters $\epsilon_{x,y}$ and $\gamma_{x,y}$. Comparing Figures \ref{fig:4a}, \ref{fig:8a} and \ref{fig:9a}, makes clear that the symmetry of the intensity distributions $\vert U^{(1)}(\mathbf{r})\vert^2$ can be manipulated by properly choosing the factorization energies $\epsilon_{x,y}$ and the parameters $\gamma_{x,y}$. Moreover, the position of the center of symmetry in the $x,y$-plane of these distributions may be controlled by changing the sign and the values of $\gamma_{x,y}$.

In contrast, the refractive index profile $n^{(2)}(\mathbf{r})$ induces a highly inhomogeneity in the optical intensity distributions $\vert U^{(2)}(\mathbf{r})\vert^{2}$, as shown in Figures \ref{fig:Modos4_SUSY} and \ref{fig:Modos5_SUSY}. More precisely, by conveniently varying the values for the parameters $\epsilon_{\nu,x}$, $\epsilon_{\nu,y}$ and $\gamma_{\nu,x}$, $\gamma_{\nu,y}$,  $\nu=1,2$, it is possible to generate displacements of the peak of intensity profile $\vert U^{(2)}(\mathbf{r})\vert^{2}$ (see Figures \ref{fig:14a} and \ref{fig:14b}). 

\section{Conclusions}
\label{Conc}

We have constructed the guided modes hosted by a sequence of optical media with refractive index distributions $n^{(j)}(\mathbf{r})$, $j=1,2,\ldots$. Such sequence arises via the supersymmetric formalism applied to the stationary paraxial Helmholtz equation for the elliptic refractive index $n^{(0)}$ \eqref{ec9} up to an arbitrary order $k$. 

The refractive profiles  $n^{(j)}(\mathbf{r})$, $j=1,2,\ldots$ so-obtained turn out to be multiparametric, as they depend on the set of parameters $\{\epsilon_{\nu,\xi},~\gamma_{\nu,\xi},~\xi=x,y,~\nu=1,2,\ldots,j\}$ that allow to tune, on demand, the spatial distribution of the refractive index in the transverse $xy$-plane. As a consequence, new optical guided modes can be generated in the {\it paraxial regime} as supersymmetric deformations of the departure GHG modes. The main characteristic of each profile $n^{(j)}(\mathbf{r})$, $j=1,2,\ldots$, is that it accommodates one additional guided mode than its predecessor $n^{(j-1)}(\mathbf{r})$, giving raise to a sequence of subsidiary modes corresponding to the sequence of effective refractive indices $\varepsilon_s^{(j)}=\epsilon_{j-s}$, $s=0,1,2\ldots,j-1$, $\left\{\epsilon_j <\epsilon_{j-1}<\cdots <\epsilon_1<\frac12\right\}$. In this sense, a more in-depth study of light propagation in these new optical media can be carried out by including subsidiary modes in the construction of wave-packets.

Furthermore, some specific selections of the parameters $\epsilon_{\nu,x}$, $\epsilon_{\nu,y}$, $\gamma_{\nu,x}$, $\gamma_{\nu,y}$, $\nu=1,2,\ldots$, generate high-asymmetric profiles. This fact opens up the possibility of designing optical materials with refractive indices that allow the propagation of light beams with specific characteristics adapted for diverse technological applications.

\section*{Acknowledgments}
This work has been partially supported by CONAHCYT Project FORDECYT-PRONACES/61533/2020, Secihti Project CBF 25-1-2875 and Instituto Polit\'ecnico Nacional Projects SIP20253949, SIP20254000, SIP20260922 and SIP20260933. SCyC is grateful to the members of the Mathematical Physics Research Group of the Valladolid University for their kind hospitality. ZGM and DOC gratefully acknowledge the financial support from SECIHTI through the postdoctoral fellowships, CVU numbers 409616 and 712322, respectively.



\begin{thebibliography}{99}

\bibitem{Gom02}
Gomez-Reino C, Perez M V and Bao C 2002 \emph{Gradient index optics} Springer-Verlag Berlin Heidelberg

\bibitem{Mar97}
Marte M A M and Stenholm S 1997 Paraxial light and atom optics: The optical Schr\"odinger equation and beyond \emph{Phys. Rev. A} \textbf{56} 2940

\bibitem{Dra99}
Dragoman D and Dragoman M 1999 Optical analogue structures to mesoscopic devices \emph{P. Quant. Elec.} \textbf{23} 131-188

\bibitem{Dra02}
Dragoman D 2002 Phase space correspondence between classical optics and quantum mechanics \emph{Progress in Optics} \emph{43} 433-496

\bibitem{Lon09}
Longhi S 2009 Quantum-optical analogies using photonic structures \emph{Light \& Photon Rev.} \textbf{3} 243-261

\bibitem{Cru15}
Cruz y Cruz S and Razo R 2015 Wave propagation in the presence of a dielectric slab: The paraxial approximation \emph{J. Phys.: Conf. Ser.} \textbf{624} 

\bibitem{Cru15a}
Cruz y Cruz S and Rosas-Ortiz O 2015 Leaky Modes of Waveguides as a Classical Optics Analogy of
Quantum Resonance \emph{Adv. Math. Phys.} \textbf{2015} 281472

\bibitem{Glo69}
Gloge D and Marcuse D 1969 Formal Quantum Theory of Light Rays \emph{J. Opt. Soc. Am.} \textbf{59} 1629-1631

\bibitem{Sto81}
Stoler D 1981 Operator methods in physical optics \emph{J. Opt. Soc. Am.} \textbf{71} 334-341

\bibitem{Nie93}
Nienhuis G and Allen L 1993 Paraxial wave optics and harmonic oscillators \emph{Phys. Rev. A} \textbf{48}  656-665

\bibitem{Cru17}
Cruz y Cruz S and Gress Z 2017 Group approach to the paraxial propagation of Hermite-Gaussian modes in a parabolic medium \emph{Ann. Phys.} \textbf{383} 257-277

\bibitem{Lon18}
Longhi S 2018 Parity-time symmetry meets photonics: A new twist in non-Hermitian optics \emph{EPL} \textbf{120} 64001

\bibitem{Cru19}
Cruz y Cruz S, Gress Z, Jim\'enez-Mac\'ias P and Rosas-Ortiz O 2019 Laguerre-Gaussian wave propagation in parabolic media \emph{Geometric Methods in Physics XXXVIII: Workshop, Bia\l owie\.za Poland} 117-128

\bibitem{Boc22}
Bocanegra I and Cruz y Cruz S 2022 Classes of balanced gain-and-loss waveguides as non-Hermtian potential hierarchies \emph{Symmetry} \textbf{14} 432

\bibitem{Cru23}
Cruz y Cruz S, Gress Z, Jim\'enez-Mac\'ias P and Rosas-Ortiz O 2023 Bessel-Gauss Beams of Arbitrary Integer Order: Propagation Profile, Coherence Properties, and Quality Factor \emph{Photonics} \textbf{10} 1162

\bibitem{Chu94}
Chumakov SM and Wolf KB 1994 Supersymmetry in Helmholtz optics \emph{Phys. Lett. A} \textbf{193} 51-53

\bibitem{Mir13}
Miri M-A, Heinrich M, El-Ganainy R and Christodoulides DN 2013 Supersymmetric optical structures \emph{Phys. Rev. Lett.} \textbf{110} 233902

\bibitem{Hei14}
Heinrich M, Miri M-A, St\"utzer S, El-Ganainy R, Nolte S, Szameit A, and Christodoulides DN 2014 Supersymmetric mode converters \emph{Nature Commun.} \textbf{5} 3698

\bibitem{Mie84}
Mielnik B 1984 Factorization method and new potentials with the oscillator spectrum \emph{J. Math. Phys.} \textbf{25} 3387-3389

\bibitem{Jun98}
Junker G and Roy P 1998 Conditionally exactly solvable potentials: A supersymmetric construction method \emph{Ann. Phys.} \textbf{270} 155-177

\bibitem{Mie00}
Mielnik B, Nieto LM and Rosas-Ortiz 2000 The finite difference algorithm for higher order supersymmetry
\emph{Phys. Lett. A} \textbf{269} 70-78

\bibitem{Coo01}
Cooper F, Khare A and Sukhatme U 2001 \emph{Supersymmetry in Quantum Mechanics} World Scientific

\bibitem{Mie04}
Mielnik B and Rosas-Ortiz O 2004 Factorization: little or great algorithm? \emph{J. Phys. A: Math. Theor.} \textbf{37} 10007

\bibitem{Suk04}
Sukumar CV 2004 Supersymmetric quantum mechanics and its applications \emph{AIP Conference Proceedings} \textbf{744} 166-235

\bibitem{Fer04}
Fern\'andez C DJ and Fern\'andez-Garc\'ia N 2004 Higher-order supersymmetric quantum mechanics \emph{AIP Conference Proceedings} \textbf{744} 236-273

\bibitem{Gan10}
Gangopadhyaya A, Mallow JV and Rasinariu C 2010 \emph{Supersymmetric Quantum Mechanics} World
Scientific

\bibitem{Inf51}
Infeld L and Hull T E 1951 The Factorization Method \emph{Rev. Mod. Phys.} \textbf{23} 21-68

\bibitem{Ber14}
Bermudez D and Fern\'andez C DJ 2014 Supersymmetric quantum mechanics and Painlev\'e equations \emph{AIP Conference Proceedings} \textbf{1575} 50-88

\bibitem{Var23}
Vargas-Cruz A, Ortiz-Campa D and D\'iaz-Bautista E 2023 Socios supersim\'etricos para el oscilador arm\'onico cu\'antico \emph{In Lee Guzm\'an E (Ed.) Memorias de la Reuni\'on Nacional Acad\'emica de F\'isica y Matem\'aticas} \textbf{28} 343-348

\bibitem{Mac18}
Macho A, Llorente R and Garc\'ia-Meca C 2018 Supersymmetric transformations in optical fibers
\emph{Phys. Rev. Appl.} \textbf{9} 014024

\bibitem{Hua22}
Huang C and Song Q 2022 Guiding flow of light with supersymmetry \emph{Light: Sci \& Appl} \textbf{11} 290

\bibitem{Zel17}
Zelaya K and Rosas-Ortiz O 2017 Exactly Solvable Time-Dependent Oscillator-Like
Potentials Generated by Darboux Transformations \emph{J. Phys.: Conf. Ser.} \textbf{839} 012018

\bibitem{Con19}
Contreras-Astorga A and Jakubsk\'y V 2019 Photonic systems with two-dimensional landscapes of complex refractive index via time-dependent supersymmetry \emph{Phys. Rev. A} \textbf{99} 053812

\bibitem{Raz19}
Razo R and Cruz y Cruz S 2019 New confining optical media generated by Darboux
transformations \emph{J. Phys.: Conf. Ser.} \textbf{1194} 012091

\bibitem{Cru20}
Cruz y Cruz S, Razo R, Rosas-Ortiz O and Zelaya K 2020 Coherent states for exactly solvable
time-dependent oscillators generated by Darboux transformations \emph{Phys. Scrip.} \textbf{95} 044009

\bibitem{Gar20}
Garc\'ia-Meca C, Macho Ortiz A, and Llorente S\'aez R 2020 Supersymmetry in the time domain and its
applications in optics \emph{Nature Commun.} \textbf{11} 813

\bibitem{Gre17}
Gress Z and Cruz y Cruz S 2017 A note on the off-axis Gaussian beams propagation in parabolic media \emph{J. Phys.: Conf. Ser} \textbf{839} 012024

\bibitem{Gre19}
Gress Z and Cruz y Cruz S 2019 Hermite coherent states for quadratic refractive index optical media \emph{In Integrability, Supersymmetry and Coherent States} Kuru S, Negro J and Nieto LM (Eds.) CRM Series in Mathematical Physics Springer: Berlin/Heidelberg Germany 323-339

\bibitem{Sie86}
Siegman AE 1986 \emph{Lasers} University Science Books 

\bibitem{Sal91}
Saleh BEA and Teich MC 1991 \emph{Fundamentals of Photonics} John Wiley and Sons: New York

\bibitem{Kot13} 
Kotlyar V V, Kovalev A A and Nalimov A G 2013 Propagation of hypergeometric
laser beams in a medium with a parabolic refractive index \emph{J. Opt.} \textbf{15} 125706

\bibitem{Pet16}
Petrov N I 2016 Spin-Dependent Transverse Force on a Vortex
Light Beam in an Inhomogeneous Medium \emph{JETP Lett.} \textbf{13} 443-448

\bibitem{Wu20}
Wu Y, Wu J, Lin Z, Fu X, Qiu H, Chen K and Deng D 2020 Propagation
properties and radiation forces of the Hermite-Gaussian vortex beam
in a medium with a parabolic refractive index \emph{Appl. Opt.} \textbf{59} 8342-8348

\bibitem{Put21}
Puttnam B J, Rademacher G and Lu\'is R S 2021 Space-division
multiplexing for optical fiber communications \emph{Optica} \textbf{8} 1186-1203

\end{thebibliography}
\end{document}